\documentclass[11pt,thmsa]{article}
\usepackage{amsmath}
\usepackage{amsfonts}
\usepackage{amssymb}
\usepackage{geometry}
\usepackage{xcolor, graphicx}
\usepackage{tikz}

\begin{document}

\title{\textbf{{\LARGE Measuring Stochastic Rationality\thanks{%
An earlier version of this paper was circulated under the title
\textquotedblleft Comparative Rationality of Random Choice
Behaviors.\textquotedblright\ We would like to dedicate this paper to the
memory of the late Peter Fishburn. Peter was an exceptional applied
mathematician who made pioneering contributions to mathematical decision
theory within economics and psychology and was one of the leading figures in
order theory at large. In fact, the present paper is firmly built on a
largely forgotten idea of Peter that dates back to 1978 on connecting
deterministic and stochastic choice theories. We will miss him dearly. We
thank Paul Feldman, John Rehbeck, Mauricio Ribeiro, Gil Riella, John Quah,
and the seminar participants at New York University and Johns Hopkins
University for their suggestions and feedback.}}}}
\author{Efe A. Ok\thanks{\linespread{1.1} \selectfont Department of
Economics and the Courant Institute of Mathematical Sciences, New York
University. E-mail: efe.ok@nyu.edu.} \and Gerelt Tserenjigmid\thanks{%
Department of Economics, the University of California Santa Cruz. Email:
gtserenj@ucsc.edu.}}
\maketitle

\begin{abstract}
\noindent Our goal is to develop a partial ordering method for comparing
stochastic choice functions on the basis of their \textit{individual
rationality}. To this end, we assign to any stochastic choice function a
one-parameter class of deterministic choice correspondences, and then check
for the rationality (in the sense of revealed preference) of these
correspondences for each parameter value. The resulting ordering detects
violations of (stochastic) transitivity as well as inconsistencies between
choices from nested menus. We obtain a parameter-free characterization and apply it to some popular stochastic
choice models. We also provide empirical applications in terms of two well-known
choice experiments.

\medskip \medskip

\noindent \textit{JEL\ Classification}: D01, D11, D81, D91.


\newpage
\end{abstract}

\section{Introduction}

Deterministic and stochastic choice theories are the two main strands of
modeling a decision maker's choice behavior. The deterministic approach is
simpler, and it provides clear criteria about when a person's choice
behavior is rational and when it is not.\footnote{%
The word \textquotedblleft rational\textquotedblright\ is a loaded one, and
many authors in economics, philosophy, and other fields assign different,
often conflicting, meanings to it. We are pragmatic about this concept. By
the \textquotedblleft rationality\textquotedblright\ of an individual's\
choice behavior, we understand that their choices in large menus are
consistent with those in pairwise choice situations; this corresponds to
preference maximizing behavior.} However, despite its time-honored stance,
this approach is hard to connect to empirical choice data and is too coarse
to model empirically relevant deviations from preference maximization in a
unified manner. In this regard, stochastic choice theory has significant
advantages. It can readily be inferred from repeated choices of a person,
and it is rich enough to capture choice anomalies.\footnote{%
On the former point, see Balakrishnan, Ok and Ortoleva (2021), and on the
latter, Gul, Natenzon, and Pesendorfer (2014), Manzini and Mariotti (2014),
Fudenberg, Iijima, and Strzalecki (2015), Aguiar (2017), Echenique, Saito,
and Tserenjigmid (2018), and Natenzon (2019), among others.} And yet, when
we model their choices as random entities, it becomes less clear when we can
classify an individual as \textquotedblleft rational,\textquotedblright\ and
how we can compare the extent of \textquotedblleft
rationality\textquotedblright\ of two individuals. For instance, if a
person's choice behavior abides by a random utility model (RUM), is this
individual necessarily rational? Or is it the case that some RUMs correspond
to \textquotedblleft more rational\textquotedblright\ behavior than other
RUMs? Which sorts of constraints do not detract from the rationality of the
classical Luce model? When we have the empirical record of the random
choices of two subjects in an experiment, how can we judge that one behaves
more rationally than another?

The goal of the present work is to introduce methods of comparing random
choice behaviors to provide concrete answers to such questions. In addition,
we would like this methodology to be computationally feasible, and easy to
implement, in the context of empirical applications. However, before we
describe our approach, we should underscore that we interpret stochastic
choice functions in this paper exclusively as models of choice for \textit{%
individual} decision-making units. This interpretation contrasts sharply
with that of discrete choice theory where such functions correspond to the
behavior of a collective whose constituents are different individuals who
make their choices in a deterministic fashion. (More on this in Section 5.1.)

The starting point of our approach is to approximate stochastic choice
functions by using a one-parameter family of deterministic choice
correspondences, which was introduced by Fishburn (1978). Put precisely, for
any stochastic choice function $\mathbb{P}$ and $\lambda \in (0,1],$ we
define $C_{\mathbb{P},\lambda }$ as the choice correspondence that chooses
in a menu $S$ all options whose choice probabilities are at least the 100$%
\times \lambda $ percent of the maximum likelihood of choice in that menu.
(So, for small enough $\lambda $, $C_{\mathbb{P},\lambda }$ selects
everything that is chosen in $S$ with positive probability, while $C_{%
\mathbb{P},1}$ selects only the items in $S$ with maximum likelihood of
being chosen.) We think of $C_{\mathbb{P},\lambda }$ as a \textit{%
deterministic approximation} of $\mathbb{P}$. There is no \textit{a priori}
reason to adopt a particular value for $\lambda $ in this approximation;
depending on the specifics of the choice environment, such as the size of
the alternative space, the duration of time one needs to make a decision,
the background noise, etc., certain values of $\lambda $ may be more
appropriate to use than others (Balakrishnan, Ok and Ortoleva (2021)). In
this paper, we do not subscribe to any particular value of $\lambda $.
Instead, we work with the whole family $\{C_{\mathbb{P},\lambda }:0<\lambda
\leq 1\}$.

As we have noted above, the \textquotedblleft rationality\textquotedblright\
of a deterministic choice correspondence is easily defined; such a
correspondence is \textit{rational} if it arises from the maximization of a
(possibly incomplete) preference relation.\footnote{%
Many experts are of the opinion that there is nothing irrational about
having incomplete preferences in general. (See, for instance, von Neumann
and Morgenstern (1947), Aumann (1962), Bewley (1986), and Schmeidler (1989),
among many others). As we will discuss later, allowing for incomplete
preferences becomes even more appealing in stochastic choice environments.}
Besides, the literature provides a variety of ways of recognizing such
correspondences by means of revealed preference postulates. Our proposal
here is to compare the rationality of two stochastic choice functions $%
\mathbb{P}$ and $\mathbb{Q}$ by comparing the rationality of the members of
the Fishburn families associated with these functions. Put precisely, we say
that $\mathbb{P}$ \textit{is at least as rational as }$\mathbb{Q}$ if $C_{%
\mathbb{P},\lambda }$ is rational for every $\lambda \in (0,1]$ for which $%
C_{\mathbb{Q},\lambda }$ is rational; we denote this situation by writing $%
\mathbb{P}\trianglerighteq _{\text{rat}}\mathbb{Q}$. When this is the case,
whichever member of the Fishburn family one uses to approximate the given
stochastic choice functions, the one associated with $\mathbb{P}$ is sure to
arise from the maximization of a preference relation if so does the one
associated with $\mathbb{Q}$. It is in this sense that $\trianglerighteq _{%
\text{rat}}$ acts as an unambiguous, albeit incomplete, ordering of
stochastic choice functions in terms of their rationality. We say that $%
\mathbb{P}$ is \textit{maximally rational} if it maximizes $\trianglerighteq
_{\text{rat}}$ across all stochastic choice functions, that is, it is at
least as rational as any stochastic choice function. In Section 3.2, we show
that such choice behavior is consistent with all but the most extreme forms
of stochastic transitivity. This witnesses the fact that our rationality
ordering is primed to detect violations of (stochastic) transitivity as well
choice inconsistencies across nested menus.

While the motivation behind the preorder $\trianglerighteq _{\text{rat}}$ is
fairly clear, and it resembles other partial ordering methods used elsewhere
in economics (such as the classical Lorenz ordering of income inequality
measurement), it is difficult to apply $\trianglerighteq _{\text{rat}}$
using its definition alone. After going through some preliminaries of
individual choice theory in Section 2, we thus obtain a parameter-free
characterization of $\trianglerighteq _{\text{rat}}$ in Section 3. This
characterization assigns to each stochastic choice function $\mathbb{P}$ a
set $\Lambda (\mathbb{P})$, and shows that $\mathbb{P}\trianglerighteq _{%
\text{rat}}\mathbb{Q}$ iff $\Lambda (\mathbb{P})\subseteq \Lambda (\mathbb{Q}%
)$. This not only makes working with $\trianglerighteq _{\text{rat}}$ easier
(as witnessed by our subsequent applications), but it also lets us define a
natural index of rationality that is consistent with $\trianglerighteq _{%
\text{rat}}$ (by using the size of $\Lambda (\mathbb{P})$). This index,
which we denote by $I_{\text{rat}}$, quantifies the extent to which one
stochastic choice function is more rational than another. Moreover, it can
be used to rank any two stochastic choice functions that happen to be
incomparable by means of $\trianglerighteq _{\text{rat}}$ (just like one can
use, say, the Gini index to compare two income distributions whose Lorenz
curves cross). 

In Section 4, we specialize our methodology to the case of stochastic choice
functions that arise from binary choice data (that is, choice from doubleton
menus). In this case, which applies to most repeated/random choice
experiments, the applicability of the order $\trianglerighteq _{\text{rat}}$
increases, and $I_{\text{rat}}$ becomes easier to compute. In this section
we also outline an algorithm to calculate $\Lambda (\mathbb{P})$ in binary
choice environments; our applications demonstrate that this algorithm is
quite fast.

With the theory developed in Sections 3 and 4, one can address the kinds of
questions we have posed above formally, and at times, obtain answers that
are not self-evident. In our first application (Section 5.1), we investigate
the \textit{individual rationality} of random utility models. Such models
arise as stochastic extensions of the classical (deterministic) rational
choice model. Perhaps for this reason, many decision theorists seem to view 
\textit{any }RUM as corresponding to the stochastic choice behavior of a
rational individual. On closer scrutiny, however, this position is suspect.
For example, while transitivity is the key property of rational
(deterministic) choice model, a RUM may fail to satisfy even the weakest
form of stochastic transitivity.\footnote{%
This is not a theoretical anomaly. In the empirical application of Section
5.4 we find that 4 (resp. 7) of the 26 subjects exhibit weakly (resp.
moderately) stochastically \textit{non}-transitive behavior that is
nonetheless consistent with the RUM model.} Similarly, while the axiomatic
basis of the former model is very much based on rationality axioms (such as
the weak axiom of revealed preference), none of the known axioms for the RUM
has the makeup of a rationality criterion.

In fact, one does not have to be an expert to see that the \textquotedblleft
rationality\textquotedblright\ of certain RUMs is not beyond reproach. Most
of us would not think of a politician as \textquotedblleft
rational\textquotedblright\ if they act as an unflinching republican, or as
an astute socialist, each with one-half probability, every time they exert
their opinion.\footnote{%
Just to be clear, the issue is not one's tastes, but the \textit{consistency 
}of them. After all, the behavior of this politician is markedly different
than acting as a republican on half of the issues and as a socialist on the
other half. Obviously, there is nothing irrational -- or random -- about the
latter type of behavior. This person, instead, takes orthogonal ideological
positions on the \textit{same} set of issues with equal probabilities. (For
instance, one day they advocate making all abortion practices illegal, and
the next day they defend allocating a large budget to open federally funded
abortion clinics, \textit{on the basis of a coin toss}.) We submit that most
of us would call this person \textquotedblleft
inconsistent,\textquotedblright\ or \textquotedblleft
irrational,\textquotedblright\ even though their behavior conforms fully
with the RUM.} As demanding as it is, the ordering $\trianglerighteq _{\text{%
rat}}$ catches this intuition accurately. Indeed, we find that a RUM may
even be \textit{minimally} rational according to $\trianglerighteq _{\text{%
rat}}$. The point is that, to warrant individual rationality, the set of
utility functions of a RUM should in some sense be coherent. In Section 5.1,
we characterize what this means for the RUM with two utility functions. In
fact, $\trianglerighteq _{\text{rat}}$ is able to rank any two such RUMs,
and classifies any such model with this consistency condition as maximally
rational. Our method is thus able to behaviorally distinguish between
various random utility models on the basis of their rationality.

A similar situation occurs with Luce-type models. First, it is readily
checked that the classical Luce model is maximally rational. On the other
hand, a general Luce model (in which the individual uses the Luce rule only
on a (consideration) subset of any given menu) may even be minimally
rational. However, we show in Section 5.2 that every 2-stage Luce rule is
maximally rational, provided it satisfies a simple consistency condition.

These observations demonstrate that if one wishes to adopt either a RUM or a
general Luce-type model in economic applications, and if they wish to
investigate the consequences of rational, yet random, choice behavior, they
need to impose suitable consistency conditions on the primitives of these
models. The preorder $\trianglerighteq _{\text{rat}},$ as well as the index $%
I_{\text{rat}}$, can be used to identify the needed consistency condition.
Even better, these measures let one perform comparative statics exercises in
which one studies how the results of an economic application alter as the
adopted model of random choice is made more rational. This is the manner in
which our theoretical contribution may be used in applied microeconomic
modeling.

Our third application pertains to the tremble model of choice (Section 5.3).
We find that no two such models with distinct tremble probabilities are
comparable by $\trianglerighteq _{\text{rat}}$. Instead, we compare the
rationality of such random choice models by using the index $I_{\text{rat}}$%
. We find that a tremble model gets more rational (with respect to $I_{\text{%
rat}}$) as the likelihood of the trembles, which are smaller than a
threshold, gets smaller. In the limit where this likelihood is zero, the
model becomes the standard utility-maximization model, which is, of course,
maximally rational. However, this is valid only for tremble probabilities
that are small enough. When these are larger than the said threshold, the
opposite conclusion is obtained. In that case, our measurement methodology
picks up the fact that there is no reason to view trembles as
\textquotedblleft mistakes\textquotedblright\ when they occur in high
frequency. When tremble probability is large enough, increasing it further
makes the model get closer to the one that chooses everything in every menu
with equal probabilities, thereby increasing its rationality (at least in
the eyes of $I_{\text{rat}}$). This makes good sense because the latter
model is maximally rational, as it simply corresponds to the behavior of a
utility-maximizer who is indifferent between all alternatives.

In Section 5.4 we provide two empirical applications. First, we use the
pooled choice data of Tversky (1969) and Regenwetter, Dana and Davis-Stober
(2011) to determine which of the involved subjects are maximally rational
according to our rationality ordering. We also make rationality comparisons
between the subjects by using the ordering $\trianglerighteq _{\text{rat}}$,
and determine the value of $I_{\text{rat}}$ for each of them. Second, we
consider the large data set of the binary choice experiments of Feldman and
Rehbeck (2022), and demonstrate the computational advantages of our
algorithm for computing $\trianglerighteq _{\text{rat}}$ and $I_{\text{rat}}$%
. In both of our empirical applications, we found that our methodology
yields novel insights about the data, and it highlights nuances that were
previously not apparent. In addition, despite its incompleteness, we
observed that $\trianglerighteq _{\text{rat}}$ is able to render a
substantial amount of comparisons in the data.

Section 6 concludes the paper with some closing remarks about future
research.

\medskip

\noindent \textsf{\textbf{Related Literature. }}The literature on measuring
the rationality of deterministic choice behavior is substantial. Afriat
(1973) studied this issue in the context of consumer demand theory, and his
work was advanced by the subsequent contributions of Varian (1990),
Echenique, Lee and Shum (2011), Dean and Martin (2016), and Aguiar and
Serrano (2017). While these papers work within specific economic
environments, there are also studies on measuring the rationality of choice
behavior in general environments; see, \textit{inter alia}, Houtman and Maks
(1985), Apesteguia and Ballester (2015), Caradonna (2020), and Ribeiro
(2021).

Other than Ribeiro (2021), none of these papers develops a partial ordering
approach, and other than Apesteguia and Ballester (2015), neither applies to
the context of stochastic choice functions. And even the powerful work of
Apesteguia and Ballester (2015), while applicable to the random choice
domain, targets predominantly the deterministic choice functions. Unlike
ours, that approach does not allow indifferences or incomparabilities, which
forces it to conclude that uniform randomization due to indifferences is the 
\textit{least} rational stochastic choice. In addition, the
Apesteguia-Ballester measure does not detect violations of stochastic
transitivity, nor is it designed to do this. More on this later.

All in all, while our broad objectives are the same, we were unable to
discern a strong connection between the present paper and the previous
literature on measuring rationality. On one hand, our approach is not
suitable for making rationality comparisons of deterministic choice
behavior. If one specializes our rationality ordering to the deterministic
context, one gets an extremely coarse ordering that renders one choice
correspondence more rational than another iff the first one is rational (in
the sense of preference maximization) and the other is not. Simply put, our
method is useless for studying the comparative rationality of deterministic
choice behaviors. On the other hand, the literature on measuring rationality
in the context of deterministic choice does not apply to the present setting
of random choice. Having said this, however, one may combine the partial
ordering method we introduce here with the deterministic methods of the
earlier literature to obtain a comprehensive approach to measuring
rationality in decision-making. We discuss this possibility in Section 3.

As surveyed by de Clippel and Rozen (2023) and Strzalecki (2023) recently,
the literature on stochastic choice is a vibrant one; about two dozen new
stochastic choice models were proposed in the past ten years. The
experimental literature documenting individuals making random choices is
also growing. It thus appears that there is a substantial need for a unified
means of comparing stochastic choice models, as well as comparing \textit{%
individual} random choice data collected in experiments. The present paper
aims to contribute to this area.

In passing, we recall that Kitamura and Stoye (2018) develops a method for
testing RUM in the classical context of consumer demand. Their method can be
extended to measure a \textquotedblleft distance\textquotedblright\ between
a given stochastic data set and the set of all RUMs in the context of
discrete choice. One may be tempted to use this distance as a rationality
index. Unfortunately, this approach would face serious computational
difficulties, as the Kitamura-Stoye approach to testing RUM is an NP-hard
problem. (As far as we know, it is not even feasible to implement for data
sets as large as in, say, Feldman and Rehbeck (2022).) Far more important is
that this approach would assume without justification that \textit{every}
RUM is fully rational. As we noted above, and will discuss further in
Section 5.1, this position is not viable.

\medskip 

\noindent \textbf{On the Source of Randomness.} There are many reasons why
people may deviate from making rational choices. These include limited
attention, status quo bias, reference-dependence, context-dependence, and
complexity, among others (de Clippel and Rozen (2023)). Most of the
rationality measures in the literature on deterministic choice do not target
a specific source of bounded rationality. Similarly, per stochastic choice
theory literature, individuals' choices are random due to several different
reasons, e.g., imperfect perceptions/attention, mistakes, and preference for
randomness, among others. In this paper, we follow the footsteps of the
literature on deterministic choice; our rationality order does not rely on a
particular source of randomness. An important advantage of this is that our
rationality order $\trianglerighteq _{\text{rat}}$ is applicable to every
model of stochastic choice. However, its conclusions can be rather coarse
for models in which there is a priori knowledge about the source of
uncertainty. In such cases, one may choose to adopt more refined methods of
rationality measurement.

\medskip 

\noindent \textbf{Computational Feasibility.} The computational complexity
of our comparison methods is encouraging. We applied the algorithm of
Section 4.2 for computing $\Lambda (\mathbb{P})$ to the experimental data of
Feldman and Rehbeck (2022). In that study the number of choice alternatives
is $27$, the number of binary menus is $79,$ and there are 144 subjects;
this is the largest random choice data set that we are aware of. On a
personal computer, our algorithm calculates $\Lambda (\mathbb{P}),$ and
hence makes rationality comparisons with $I_{\text{rat}}$, in the case of
this data within minutes.

This issue has begun to receive attention only recently in the economics
literature, but is nevertheless quite important, especially with large data
sets becoming available through online experiments and digital choice
observations made in the marketplace. Measuring the deviations from
specific stochastic choice models may be a daunting task with such large
data sets. For instance, even in the case of binary menus, we do not know
any computationally feasible method to measure deviations from the RUM when
the number of menus is large.

\section{Rationality and Choice}

\noindent Throughout this paper, unless otherwise stated explicitly, $X$
stands for a nonempty finite set with $\left\vert X\right\vert \geq 3$, and $%
\mathfrak{X}$ denotes the collection of all nonempty finite subsets of $X.$

\subsection{Deterministic Choice}

\noindent \textsf{\textbf{Preference Relations. }}A \textbf{preorder} $%
\succsim $ on $X$ is a reflexive and transitive binary relation on $X.$ (The
asymmetric part of $\succsim $ is denoted by $\succ ,$ and its symmetric
part by $\sim $.) For any $S\in \mathfrak{X},$ we denote the set of all $%
\succsim $\textbf{-maximal elements} in $S$ by \textrm{{\footnotesize 
\textbf{MAX}}}$(S,\succsim ),$ that is, $x\in $ \textrm{{\footnotesize 
\textbf{MAX}}}$(S,\succsim )$ iff $x\in S$ and there is no $y\in S$ with $%
y\succ x.$ If $\succsim $ is total as well, that is, either $x\succsim y$ or 
$y\succsim x$ for every $x,y\in X,$ we say that $\succsim $ is a \textbf{%
complete preorder} on $X.$ In that case, an element $x$ of $S$ is $\succsim $%
-maximal iff it is $\succsim $-maximum, that is, \textrm{{\footnotesize 
\textbf{MAX}}}$(S,\succsim )=\max (S,\succsim ):=\{x\in S:x\succsim y$ for
all $y\in S\}.$

Throughout this paper, by a \textbf{preference relation} on $X$, we simply
mean a preorder on $X.$ Thus, in our usage of the word, a preference
relation is always transitive, but it need not be complete.

\bigskip 

\noindent \textsf{\textbf{Choice Correspondences. }}By a \textbf{choice
correspondence} on $\mathfrak{X},$ we mean a set-valued map $C:\mathfrak{X}%
\rightarrow \mathfrak{X}$ such that $C(S)\subseteq S.$ For any such function 
$C$ and any $S\in \mathfrak{X},$ the set $C(S)$ is interpreted as the set of
all feasible alternatives in $S$ that the agent (whose choice behavior is
modeled by $C$) deems worthy of choice.

\bigskip

\noindent \textsf{\textbf{Rationality of Choice Correspondences. }}We think
of the choices of a person as \textquotedblleft rational\textquotedblright\
if in any given feasible menu, this individual deems choosable only those
items that are not dominated\ according to some consistent ranking. By the
\textquotedblleft consistency\textquotedblright\ of this ranking, we mean
its transitivity, so this way of looking at things suggests qualifying a
choice correspondence $C$ on $X$ as \textquotedblleft
rational\textquotedblright\ if there is a preorder $\succsim $ on $X$ such
that $C(S)$ equals the set of all $\succsim $-maximal elements in $S,$ for
every $S\in \mathfrak{X}$. Thus:

\bigskip

\noindent \textsc{Definition.}\textit{\textbf{\ }}Let $C$ be a choice
correspondence on $\mathfrak{X}.$ We say that $C$ is\textbf{\ rational} if
there is a preorder $\succsim $ on $X$ such that $C=$ {\footnotesize \textbf{%
MAX}}$(\cdot ,\succsim ).$ We refer to $\succsim $ as a \textbf{preference
relation that rationalizes} $C$.\footnote{%
This relation is not unique, but among all that rationalizes $C,$ there is a
unique preference relation whose indifference part matches the behavioral
indifference declarations of $C;$ see Eliaz and Ok (2006). Alternatively,
one can allow for choice deferrals (i.e., permit $C$ be empty-valued) to get
at this issue; see Gerasimou (2018).}

\bigskip

In passing, we would like to emphasize that we do not require a preference
relation that rationalizes a choice correspondence be complete. The
completeness of a preference relation is about being \textit{decisive},
which is a behavioral notion that is distinct from \textit{deciding
consistently}. After all, a rational person may well be unable to compare
two options due to insufficient information, or perception difficulties (as
in just-noticeable differences), or due to the multiplicity of attributes
relevant to choice. (As a matter of fact, this may well be the reason for
one's randomized choice behavior; see Ok and Tserenjigmid (2022).) We will
not discuss this point any further here. Many authors, such as Aumann,
Bewley, Luce, Raiffa, and Schmeidler, have written extensively on this point,
and there is now quite a substantive literature in decision theory on
rational decision making with incomplete preferences. For an overall
discussion of the matter and references, we refer the reader to Eliaz and
Ok (2006).

In the context of the present work, the problems with insisting on the
completeness of the rationalizing preferences go beyond these general
considerations. Indeed, as we elaborate toward the end of Section 3, doing
this would result in undesirable consequences for the measurement
methodology we develop in this paper.

\bigskip

\noindent \textsf{\textbf{Characterization of Rational Choice
Correspondences. }}There are various ways of characterizing rational choice
correspondences. We adopt the approach of Ribeiro and Riella (2017) which is
based on the following three simple properties imposed on $C$:

\bigskip

\noindent \textbf{The Chernoff Axiom.} For every $S,T\in \mathfrak{X}$ with $%
S\subseteq T,$ we have $C(T)\cap S\subseteq C(S).$

\medskip

\noindent \textbf{The Condorcet Axiom. }For every $S\in \mathfrak{X}$ and $%
x\in S,$ if $x\in C\{x,y\}$ for all $y\in S,$ then $x\in C(S).$

\medskip

\noindent \textbf{The No-Cycle Axiom. }For every $x,y,z\in X,$ $%
\{x\}=C\{x,y\}$ and $\{y\}=C\{y,z\}$ imply $\{x\}=C\{x,z\}.$

\bigskip

There is no need to discuss the interpretation of these properties here.
They are entirely standard, and characterize the rational choice
correspondences:

\bigskip

\noindent \textsc{Theorem 2.2.} [Ribeiro and Riella, 2017] \textit{A choice
correspondence on $\mathfrak{X}$\textsl{\ }is rational if, and only if, it
satisfies the Chernoff, Condorcet, and No-Cycle Axioms.}

\subsection{Stochastic Choice}

\noindent \textsf{\textbf{Stochastic Choice Functions.}} By a \textbf{%
stochastic choice function} on $\mathfrak{X},$ we mean a function $\mathbb{P}%
:X\times \mathfrak{X}\rightarrow \lbrack 0,1]$ such that%
\begin{equation*}
\sum_{x\in S}\mathbb{P}(x,S)=1\hspace{0.2in}\text{and}\hspace{0.2in}\mathbb{P%
}(y,S)=0
\end{equation*}%
for every $S\in \mathfrak{X}$ and $y\in X\backslash S.$ The collection of
all such functions is denoted by \textsf{\textbf{scf}}$(X).$

For any $\mathbb{P}\in $ \textsf{\textbf{scf}}$(X)$ and $S\in \mathfrak{X}$,
the map $x\mapsto \mathbb{P}(x,S)$ defines a probability distribution on $S.$
In this paper, we look at these distributions from the individualistic
perspective (as opposed to that of the literature on discrete choice). For
instance, we may imagine that a person has been observed to make choices
from the feasible menu $S$ multiple times, and $\mathbb{P}(x,S)$ is simply
the relative frequency of the times that she has chosen $x$ from $S.$ This
is the \textit{empirical} interpretation of $\mathbb{P},$ and is very much
in line with the classical revealed preference theory. Alternatively, one
can suppose that the choice behavior of the individual is intrinsically
probabilistic, or that their choice behavior is deterministic over $S$, but
the modeler has only partial information about it. As such, $\mathbb{P}(x,S)$
is considered as the actual probability of the agent choosing $x$ from $S;$
this is the \textit{theoretical} interpretation of $\mathbb{P}$.

\bigskip

\noindent $\lambda $\textsf{\textbf{-Choice Correspondences.}} We now turn
to the issue of deducing a deterministic representation for a given
stochastic choice function. This is the first step towards the development
of our method of measuring stochastic rationality.

For any $\mathbb{P}\in $ \textsf{\textbf{scf}}$(X),$ we define the map $%
\mathbb{P}^{\ast }:X\times \mathfrak{X}\rightarrow \lbrack 0,1]$ by%
\begin{equation}
\mathbb{P}^{\ast }(x,S):=\frac{\mathbb{P}(x,S)}{\underset{\omega \in S}{\max 
}\text{ }\mathbb{P}(\omega ,S)}.  \label{star}
\end{equation}%
That is, $\mathbb{P}^{\ast }(x,S)$ is the ratio of the likelihood of $x$
being chosen in the menu $S$ to that of an alternative in $S$ with the
maximum such likelihood. Thus, for instance, if $\mathbb{P}^{\ast }(x,S)<%
\frac{1}{10},$ we understand that there is an alternative in $S$ that is 10
times more likely to be chosen than $x$.

Next, for any $\mathbb{P}\in $ \textsf{\textbf{scf}}$(X)$ and $\lambda \in
\lbrack 0,1]$, we define the $\lambda $\textbf{-choice correspondence
induced by} $\mathbb{P}$ as the map $C_{\mathbb{P},\lambda }:\mathfrak{X}%
\rightarrow \mathfrak{X}$ with%
\begin{equation*}
C_{\mathbb{P},\lambda }(S):=\left\{ x\in S:\mathbb{P}^{\ast }(x,S)\geq
\lambda \right\} 
\end{equation*}%
if $0<\lambda \leq 1,$ and with%
\begin{equation*}
C_{\mathbb{P},\lambda }(S):=\left\{ x\in S:\mathbb{P}(x,S)>0\right\} 
\end{equation*}%
if $\lambda =0.$ We refer to $\{C_{\mathbb{P},\lambda }:0\leq \lambda \leq
1\}$ as the \textbf{Fishburn family} \textbf{associated with} $\mathbb{P}$.$%
\footnote{$\lambda $-choice correspondences induced by a $\mathbb{P}\in $ 
\textsf{\textbf{scf}}$(X)$ were first considered by Fishburn (1978) who
sought the characterization of $\mathbb{P}$ such that for every $\lambda \in
\lbrack 0,1]$, there is a (utility) function $u_{\lambda }:X\rightarrow 
\mathbb{R}$ with $C_{\mathbb{P},\lambda }(S)=\arg \max u_{\lambda }(S)$ for
every $S\in \mathfrak{X}$. Ok and Tserenjigmid (2022) introduce methods of
identifying if, and when, a stochastic choice model may be thought of as
arising due to indifference, incomparability, or experimentation, by
studying properties of choice correspondences $C_{\mathbb{P},0}$ and $C_{%
\mathbb{P},1}$.}$ Balakrishnan, Ok and Ortoleva (2021) have recently used
this family to \textquotedblleft infer\textquotedblright\ a deterministic
choice correspondence from $\mathbb{P}$, and provided an axiomatic
characterization for it.\footnote{%
For any $\lambda ,$ one can also characterize $C_{\mathbb{P},\lambda }$ as
the nearest (deterministic) choice correspondence to $\mathbb{P}$ relative
to a suitable metric on \textsf{\textbf{scf}}$(X)$. However, this is a
technical tangent for our present purposes, so we do not pursue it in this
paper.}

Intuitively speaking, our goal here is to approximate a given stochastic
choice function $\mathbb{P}$ on $\mathfrak{X}$ by means of a deterministic
choice correspondence. For this approximation to be meaningful from the
behavioral perspective, it is desirable that we use a choice correspondence $%
C_{\mathbb{P}}$ so that, for any menu $S,$ the set $C_{\mathbb{P}}(S)$
consists of all items in $S$ that have a \textquotedblleft
significant\textquotedblright\ probability of being chosen in $S$,
eliminating, for instance, items that are chosen by mistake, or in a rush,
etc.. Unfortunately, there is no unique best way of doing this. If $\mathbb{P%
}(x,\{x,y\})=\frac{99}{100},$ it may make good sense to use $C_{\mathbb{P}%
,1}\{x,y\}=\{x\}$ to represent the choices of the agent deterministically,
but if $\mathbb{P}(x,\{x,y\})=\frac{51}{100},$ then using $C_{\mathbb{P},1}$
would hardly be reasonable; in that case it seems more meaningful to use $C_{%
\mathbb{P},0}\{x,y\}=\{x,y\}$ to this end. Conceptually speaking, this is
why Fishburn was forced to consider a multitude of approximations.

In words, $C_{\mathbb{P},\lambda }(S)$ consists of all feasible items in $S$
for which there are no alternatives that are $\frac{1}{\lambda }$ times more
likely to be chosen in $S.$ For $\lambda $ close to 1, it seems
unexceptionable that we qualify the members of $C_{\mathbb{P},\lambda }(S)$
as the potential \textquotedblleft choices\textquotedblright\ of the agent
from $S,$ and conversely, for $\lambda $ close to 0, it stands to reason to
think of the members of $S\backslash C_{\mathbb{P},\lambda }(S)$ as the
objects that are chosen only by mistake. In this paper, we stay away from
the issue of choosing an appropriate $\lambda $. Instead, we adopt a partial
ordering approach in which \textit{all }$C_{\mathbb{P},\lambda }$s are used
jointly.

In passing, we note that $\lambda \mapsto C_{\mathbb{P},\lambda }$ is a
decreasing map on $[0,1],$ that is,%
\begin{equation*}
\alpha >\beta \hspace{0.2in}\text{implies\hspace{0.2in}}C_{\mathbb{P},\alpha
}(S)\subseteq C_{\mathbb{P},\beta }(S)\text{ for every }S\in \mathfrak{X}%
\text{.}
\end{equation*}%
In particular,%
\begin{equation*}
C_{\mathbb{P},1}(S)\subseteq C_{\mathbb{P},\lambda }(S)\subseteq C_{\mathbb{P%
},0}(S)
\end{equation*}%
for every $S\in \mathfrak{X}$ and $\lambda \in \lbrack 0,1].$ Moreover, $%
\lambda \mapsto C_{\mathbb{P},\lambda }$ is continuous at 0 (regardless of
how one may topologize the codomain of this map). Indeed, since $X$ is
finite, $\lambda (\mathbb{P)}:=\min \mathbb{P}^{\ast }(x,S),$ where the
minimum is taken over all $(x,S)\in X\times \mathfrak{X}$ with $\mathbb{P}%
(x,S)>0$, is a (well-defined) positive number, and $C_{\mathbb{P},0}=C_{%
\mathbb{P},\lambda }$ for every $0\leq \lambda \leq \lambda (\mathbb{P}).$
In particular, any statement about $C_{\mathbb{P},\lambda }$ that holds
\textquotedblleft for all $\lambda \in (0,1]$\textquotedblright\ holds
\textquotedblleft for all $\lambda \in \lbrack 0,1].$\textquotedblright

\bigskip

\noindent $\lambda $\textsf{\textbf{-Rational Stochastic Choice Functions}}.
For any fixed $\lambda ,$ we think of $C_{\mathbb{P},\lambda }$ as a
deterministic choice correspondence that approximates $\mathbb{P}$ in a
behavioral sense. Insofar as this deterministic approximation is viewed
acceptable, it makes sense to qualify $\mathbb{P}$ as \textquotedblleft
rational\textquotedblright\ provided that $C_{\mathbb{P},\lambda }$ is
rational. This prompts:

\bigskip

\noindent \textsc{Definition.}\textit{\textbf{\ }}For any $\mathbb{P}\in $ 
\textsf{\textbf{scf}}$(X)$ and $\lambda \in \lbrack 0,1],$ we say that $%
\mathbb{P}$ is $\lambda $\textbf{-rational} if $C_{\mathbb{P},\lambda }$ is
rational.

\bigskip

In view of Theorem 2.2, and the definition of $C_{\mathbb{P},\lambda },$ the
notion of $\lambda $-rationality is readily characterized by the following
behavioral conditions on $\mathbb{P}$:

\bigskip

\noindent $\mathcal{\lambda }$\textbf{-Stochastic Chernoff Axiom.} For every 
$S,T\in \mathfrak{X}$ with $S\subseteq T,$ and $x\in S,$%
\begin{equation*}
\mathbb{P}^{\ast }(x,T)\geq \lambda \hspace{0.2in}\text{implies\hspace{0.2in}%
}\mathbb{P}^{\ast }(x,S)\geq \lambda .
\end{equation*}

\noindent $\mathcal{\lambda }$\textbf{-Stochastic Condorcet Axiom. }For
every $S\in \mathfrak{X}$ and $x\in S,$ we have $\mathbb{P}^{\ast }(x,S)\geq
\lambda $ whenever $\mathbb{P}^{\ast }(x,\{x,y\})\geq \lambda \ $for every $%
y\in S.$

\medskip

\noindent $\mathcal{\lambda }$\textbf{-Stochastic Transitivity.\ }For every $%
x,y,z\in X,$%
\begin{equation*}
\mathbb{P}^{\ast }(y,\{x,y\})<\lambda \hspace{0.1in}\text{ and } \hspace{%
0.1in}\mathbb{P}^{\ast }(z,\{y,z\})<\lambda \hspace{0.2in}\text{ imply } 
\hspace{0.2in}\mathbb{P}(z,\{x,z\})^{\ast }<\lambda.
\end{equation*}

\bigskip

It is worth noting that the first two of these properties demand the random
choices be consistent across nested menus. In contrast, the third property
applies only to pairwise choice situations, and imposes a sense in which the
choice frequencies behave transitively depending on the parameter $\lambda .$
For $\lambda =1,$ this property is none other than the (strict version) of 
\textit{weak stochastic transitivity}, the most common notion of
transitivity used for stochastic choice functions.

It is easily verified that $\mathbb{P}$ satisfies the $\lambda $-stochastic
Chernoff Axiom iff $C_{\mathbb{P},\lambda }$ satisfies the Chernoff Axiom.
Similarly, $\mathbb{P}$ satisfies the $\lambda $-stochastic Condorcet
(resp., Transitivity) Axiom iff $C_{\mathbb{P},\lambda }$ satisfies the
Condorcet (resp., No-Cycle) Axiom. (We omit the straightforward proofs.)
Consequently, an immediate application of Theorem 2.2 yields:

\bigskip

\noindent \textsc{Theorem 2.3.} \textit{For any $\lambda \in (0,1],$ a
stochastic choice function on }$\mathfrak{X}$\textit{\ is $\lambda $%
-rational if, and only if, it satisfies the $\lambda $-Stochastic Chernoff,
Condorcet and Transitivity Axioms.}

\bigskip

It is important to note at the outset that the notion of $\lambda $%
-rationality is not monotonic in $\lambda $. As a first application of
Theorem 2.3, we provide a demonstration of this.

\bigskip 

\noindent \textsc{Example 2.1}. Where $X:=\{x,y,z\}$, let $\mathbb{P}$ be
the stochastic choice function on $\mathfrak{X}$ with 
\begin{equation*}
\mathbb{P}(x,\{x,y\})=\tfrac{4}{5},\,\mathbb{P}(y,\{y,z\})=\mathbb{P}%
(z,\{x,z\})=\tfrac{2}{3},\text{ and }\mathbb{P}(x,X)=\mathbb{P}(z,X)=\tfrac{6%
}{13}.
\end{equation*}%
The set of $\lambda $ for which $\mathbb{P}$ is $\lambda $-rational is $(0,%
\frac{1}{6}]\cup (\frac{1}{4},\frac{1}{2}]$. Hence, $\mathbb{P}$ is $0$%
-rational and $\frac{1}{2}$-rational, but not $\frac{1}{5}$-rational.
Furthermore, when $\lambda \in (\frac{1}{2},1]$, $\mathbb{P}$ does not obey
the $\lambda -$Stochastic Transitivity Axiom, capturing the fact the weakest
form of stochastic transitivity is violated. Moreover, when $\lambda \in (%
\frac{1}{6},\frac{1}{4}]$, $\mathbb{P}$ does not obey the $\lambda -$%
Stochastic Condorcet Axiom, capturing inconsistencies in choices across
nested menus.\thinspace\ $\square $

\section{Comparative Rationality of Random Choice}

\subsection{The General Case}

\noindent \textsf{\textbf{The Comparative Rationality Ordering. }}The
following definition introduces the central object of the present study.

\bigskip

\noindent \textsc{Definition.}\textit{\textbf{\ }}The \textbf{comparative
rationality ordering }is the binary relation $\trianglerighteq _{\text{rat}}$
on \textsf{\textbf{scf}}$(X)$ with 
\begin{equation}
\mathbb{P}\trianglerighteq _{\text{rat}}\mathbb{Q\hspace{0.2in}}\text{iff%
\hspace{0.2in}}\mathbb{P}\text{ is }\lambda \text{-rational for every }%
\lambda \in (0,1]\text{ for which }\mathbb{Q}\text{ is }\lambda \text{%
-rational.}  \label{ratt}
\end{equation}%
When $\mathbb{P}\trianglerighteq _{\text{rat}}\mathbb{Q},$ we say that $%
\mathbb{P}$ is \textbf{at least as rational} as $\mathbb{Q}.$

\bigskip

It is readily checked that $\trianglerighteq _{\text{rat}}$ is a preorder on 
\textsf{\textbf{scf}}$(X)$. The following example shows that there are
maximum and minimum elements in \textsf{\textbf{scf}}$(X)$ with respect to
this preorder.

\bigskip

\noindent \textsc{Example 3.1}. Let $\mathbb{P}$ be a $\{0,1\}$-valued
stochastic choice function on $\mathfrak{X}$. Then, $C_{\mathbb{P}\text{,}%
\lambda }=C_{\mathbb{P}\text{,}1}$ for every $\lambda \in (0,1]$, so it
follows that $\mathbb{P}$ is $\lambda $-rational for every $\lambda \in (0,1]
$ iff $C_{\mathbb{P}\text{,}1}$ is rational. Thus, any $\mathbb{P}\in $ 
\textsf{\textbf{scf}}$(X)$ for which $C_{\mathbb{P}\text{,}1}$ is rational
is at least as rational as any stochastic choice function on $\mathfrak{X}.$
Conversely, any $\mathbb{P}\in $ \textsf{\textbf{scf}}$(X)$ for which $C_{%
\mathbb{P}\text{,}1}$ is not rational is not $\lambda $-rational for any $%
\lambda \in \lbrack 0,1],$ so $\mathbb{Q}\trianglerighteq _{\text{rat}}%
\mathbb{P}$ holds for all $\mathbb{Q}\in $ \textsf{\textbf{scf}}$(X)$.%
\footnote{\linespread{1.1} \selectfont The stochastic choice rules
considered in this example prove that $\max ($\textsf{\textbf{scf}}$%
(X),\trianglerighteq _{\text{rat}})$ and $\min ($\textsf{\textbf{scf}}$%
(X),\trianglerighteq _{\text{rat}})$ are nonempty sets, but they are
admittedly artificial. We will encounter more interesting examples of
stochastic choice functions that belong to these sets in Section 5.} $%
\square $

\bigskip

\noindent \textsf{\textbf{Characterization of }}$\trianglerighteq _{\text{rat%
}}$. While the motivation behind $\trianglerighteq _{\text{rat}}$ is fairly
straightforward, this preorder is not easy to apply due to its dependence on
a continuous parameter, namely, $\lambda .$ As a first order of business,
then, we search for a parameter-free method of checking for the
applicability of $\trianglerighteq _{\text{rat}}$. To this end, for any
given $\mathbb{P}\in $ \textsf{\textbf{scf}}$(X),$ we define the following
three sets of real numbers:%
\begin{equation*}
\text{Ch}(\mathbb{P}):=\bigcup\limits_{\substack{ S,T\in \mathfrak{X}  \\ %
S\subseteq T  \\ x\in S}}\left( \mathbb{P}^{\ast }(x,S),\mathbb{P}^{\ast
}(x,T)\right] ,\hspace{0.2in}\text{Con}(\mathbb{P}):=\bigcup\limits 
_{\substack{ S\in \mathfrak{X}  \\ x\in S}}(\mathbb{P}^{\ast }(x,S),\underset%
{y\in S}{\min }\,\mathbb{P}^{\ast }(x,\{x,y\})],
\end{equation*}%
and%
\begin{equation}
\text{ST}(\mathbb{P}):=\bigcup\limits_{x,y,z\in X}(\max \{\mathbb{P}^{\ast
}(y,\{x,y\}),\mathbb{P}^{\ast }(z,\{y,z\})\},\mathbb{P}^{\ast }(z,\{x,z\})]%
\text{.}  \label{st}
\end{equation}%
We refer to these as the \textbf{Chernoff}, \textbf{Condorcet} and \textbf{%
Stochastic Transitivity} sets of $\mathbb{P}$, respectively. Finally, we
define%
\begin{equation*}
\Lambda (\mathbb{P}):=\text{Ch}(\mathbb{P})\cup \text{Con}(\mathbb{P})\cup 
\text{ST}(\mathbb{P}).
\end{equation*}

For any $\lambda \in (0,1],$ Theorem 2.3 says that $\mathbb{P}$ is not $%
\lambda $-rational iff it violates at least one of the $\lambda $-Stochastic
Chernoff, Condorcet and Transitivity Axioms. In turn, it is easily checked
that $\mathbb{P}$ violates the $\lambda $-Stochastic Chernoff Axiom iff $%
\lambda \in $ Ch$(\mathbb{P}).$ Similarly, $\mathbb{P}$ violates the $%
\lambda $-Stochastic Condorcet (resp., Transitivity) Axioms iff $\lambda \in 
$ Con$(\mathbb{P})$ (resp., $\lambda \in $ ST$(\mathbb{P})$). Therefore, 
\begin{equation*}
\Lambda (\mathbb{P})=\{\lambda \in (0,1]:\mathbb{P}\text{ is not }\lambda 
\text{-rational}\},
\end{equation*}%
and the following obtains as an immediate consequence of this observation:

\bigskip

\noindent \textsc{Theorem 3.1.}\textit{\textbf{\ }For any\textsl{\ }$\mathbb{%
P}$ and $\mathbb{Q}$ in \textsf{\textbf{scf}}$(X)$,%
\begin{equation*}
\mathbb{P}\trianglerighteq _{\text{rat}}\mathbb{Q\hspace{0.2in}}\text{if and
only if}\mathbb{\hspace{0.2in}}\Lambda (\mathbb{P})\subseteq \Lambda (%
\mathbb{Q}).
\end{equation*}%
}This is a parameter-free characterization of our comparative rationality
ordering. We will work with this characterization in the rest of the paper,
but here is an immediate illustration of its use.

\bigskip

\noindent \textsc{Example 3.2}. Consider the stochastic choice function $%
\mathbb{P}$ on $\mathfrak{X}$ with $\mathbb{P}(x,S):=\frac{1}{\left\vert
S\right\vert }$ for every $S\in \mathfrak{X}$ and $x\in S$. Then, $\mathbb{P}%
^{\ast }(x,S)=1$ for every $S\in \mathfrak{X}$ and $x\in S$, whence the sets
Ch$(\mathbb{P})$, Con$(\mathbb{P})$ and ST$(\mathbb{P})$ are all empty. It
follows from Theorem 3.1 that $\mathbb{P}$ is at least as rational as any
stochastic choice function on $\mathfrak{X}$. $\square $

\bigskip

\noindent \textsf{\textbf{An Index of Rationality. }}The rationality order $%
\trianglerighteq _{\text{rat}}$ is a dominance ordering, and as such, it is
quite demanding. Indeed, we will later see instances in which $%
\trianglerighteq _{\text{rat}}$ fails to rank distinct stochastic choice
functions. Moreover, even when $\mathbb{P}\trianglerighteq _{\text{rat}}%
\mathbb{Q},$ we have presently no way of quantifying the strength of the
dominance of $\mathbb{P}$ over $\mathbb{Q}$. Fortunately, Theorem 3.1
suggests a way of dealing with both of these problems at one stroke by
comparing the sizes of $\Lambda (\mathbb{P})$ and $\Lambda (\mathbb{Q}).$
Since $\Lambda (\mathbb{P})$ is the union of finitely many subintervals of $%
(0,1],$ it is a Borel set. We can thus use the Lebesgue measure Leb$(\cdot )$
to this end, and define a rationality index $I_{\text{rat}}:$ \textsf{%
\textbf{scf}}$(X)\rightarrow \lbrack 0,1]$ by%
\begin{equation*}
I_{\text{rat}}(\mathbb{P}):=1-\text{Leb}(\Lambda (\mathbb{P})).
\end{equation*}%
By Theorem 3.1, this index is consistent with $\trianglerighteq _{\text{rat}}
$. Moreover, as we shall see shortly, $I_{\text{rat}}(\mathbb{P})=1$ iff $%
\mathbb{P}$ is at least as rational as any member of \textsf{\textbf{scf}}$%
(X)$.

\bigskip 

\noindent \textsc{Example 2.1 }[\textsc{continued}]. For the stochastic
choice function of Example 2.1, we have Ch$(\mathbb{P})=\varnothing ,$ Con$(%
\mathbb{P})=(\frac{1}{6},\frac{1}{4}]$ and ST$(\mathbb{P})=(\frac{1}{2},1].$
Thus, $\Lambda (\mathbb{P})=(\frac{1}{6},\frac{1}{4}]\cup (\frac{1}{2},1]$
and $I_{\text{rat}}(\mathbb{P})=\frac{5}{12}$.

\subsection{Maximal Rationality}

\noindent \textsf{\textbf{Maximal and Minimal Rationality. }}We say that a
stochastic choice function $\mathbb{P}$ on $\mathfrak{X}$ is \textbf{%
maximally rational} if $\mathbb{P}$ is at least as rational as any $\mathbb{Q%
}$ in \textsf{\textbf{scf}}$(X)$, that is, 
\begin{equation*}
\mathbb{P}\trianglerighteq _{\text{rat}}\mathbb{Q}\text{\hspace{0.2in}for
every }\mathbb{Q}\in \text{\textsf{\textbf{scf}}}(X).
\end{equation*}%
Obviously, if $\mathbb{P}$ is $\lambda $-rational for every $\lambda \in
\lbrack 0,1],$ then it is maximally rational. The converse of this is also
true, because, as we have already seen above, there are stochastic choice
functions on $\mathfrak{X}$ that are $\lambda $-rational for every $\lambda
\in \lbrack 0,1].$ Put differently, $\mathbb{P}$ is maximally rational iff $%
\Lambda (\mathbb{P})=\varnothing $. In turn, the latter condition entails
trivially that $I_{\text{rat}}(\mathbb{P})=1.$ Conversely, since $\Lambda (%
\mathbb{P})$ is the union of finitely many subintervals of $(0,1],$ if it is
nonempty, it must contain a nondegenerate interval, which means $I_{\text{rat%
}}(\mathbb{P})<1.$ Thus:

\bigskip

\noindent \textsc{Proposition 3.2.}\textit{\textbf{\ }For any\textsl{\ }$%
\mathbb{P}$ in \textsf{\textbf{scf}}$(X)$, the following are equivalent}:%
\textit{\ }

\textit{a. $\mathbb{P}$ is maximally rational};\textit{\ }

\textit{b. $\mathbb{P}$ is $\lambda $-rational for every $\lambda \in (0,1]$}%
;\textit{\ }

\textit{c. }$I_{\text{rat}}(\mathbb{P})=1.$

\bigskip

Dually, we say that $\mathbb{P}\in $ \textsf{\textbf{scf}}$(X)$ is \textbf{%
minimally rational} if $\mathbb{Q}\trianglerighteq _{\text{rat}}\mathbb{P}$
for every $\mathbb{Q}\in $ \textsf{\textbf{scf}}$(X).$ This is equivalent to
say that $\mathbb{P}$ fails to be $\lambda $-rational for any $\lambda \in
\lbrack 0,1],$ and that $I_{\text{rat}}(\mathbb{P})=0.$

\bigskip

\noindent \textsf{\textbf{Notions of Stochastic Transitivity. }}Transitivity
is a key property of rationality. In the context of deterministic choice,
this property pertains to revealed preference relations, and it is derived
from basic rationality postulates (such as WARP). In the context of
stochastic choice, various versions of this property are formulated directly
in terms of stochastic choice functions. Abbreviating the term
\textquotedblleft stochastic transitivity\textquotedblright\ to
\textquotedblleft s-transitivity\textquotedblright\ for brevity, we next
recall the definitions of these properties.

A stochastic choice function $\mathbb{P}$ on $\mathfrak{X}$ is said to be 
\textbf{weakly s-transitive }if $\mathbb{P}(x,\{x,y\})\geq \tfrac{1}{2}$ and 
$\mathbb{P}(y,\{y,z\})\geq \tfrac{1}{2}$ imply%
\begin{equation}
\mathbb{P}(x,\{x,z\})\geq \tfrac{1}{2}  \label{staa}
\end{equation}%
for every $x,y,z\in X,$ and \textbf{almost weakly s-transitive} if (\ref%
{staa}) holds for every $x,y,z\in X$ with $\mathbb{P}(x,\{x,y\})>\tfrac{1}{2}
$ and $\mathbb{P}(y,\{y,z\})>\tfrac{1}{2}$. In turn, $\mathbb{P}$ is called 
\textbf{moderately s-transitive }if $\mathbb{P}(x,\{x,y\})\geq \tfrac{1}{2}$
and $\mathbb{P}(y,\{y,z\})\geq \tfrac{1}{2}$ imply%
\begin{equation}
\mathbb{P}(x,\{x,z\})\geq \min \{\mathbb{P}(x,\{x,y\}),\mathbb{P}%
(y,\{y,z\})\}  \label{sta}
\end{equation}%
for every $x,y,z\in X$, while we say that it is \textbf{almost moderately
s-transitive} if (\ref{sta}) holds for every $x,y,z\in X$ with $\mathbb{P}%
(x,\{x,y\})>\tfrac{1}{2}$ and $\mathbb{P}(y,\{y,z\})>\tfrac{1}{2}$. It is
plain that \textquotedblleft almost\textquotedblright\ versions of weak and
moderate s-transitivity are minor weakenings of these stochastic
transitivity concepts. Finally, $\mathbb{P}$ is said to be \textbf{strongly
s-transitive }if $\mathbb{P}(x,\{x,y\})\geq \tfrac{1}{2}$ and $\mathbb{P}%
(y,\{y,z\})\geq \tfrac{1}{2}$ imply%
\begin{equation*}
\mathbb{P}(x,\{x,z\})\geq \max \{\mathbb{P}(x,\{x,y\}),\mathbb{P}%
(y,\{y,z\})\}
\end{equation*}%
for every $x,y,z\in X$.

\bigskip

\noindent \textsf{\textbf{Maximal Rationality and Stochastic Transitivity. }}%
Since every maximally rational $\mathbb{P}\in $ \textsf{\textbf{scf}}$(X)$
is 1-rational, every such stochastic choice function is almost weakly
s-transitive. Less trivially, the following stronger result is true.

\bigskip

\noindent \textsc{Proposition 3.3.}\textit{\textbf{\ }Every maximally
rational} $\mathbb{P}\in $ \textsf{\textbf{scf}}$(X)$\textit{\ is almost
moderately s-transitive.}

\smallskip 

\noindent \textsc{Proof. }See Appendix.

\bigskip

This observation corroborates the idea that the rationality ordering $%
\trianglerighteq _{\text{rat}}$ detects violations of stochastic
transitivity. Apparently, no stochastic choice function that fails almost
moderate s-transitivity, let alone almost weak s-transitivity, can be
maximally rational. It readily follows from this observation that not every
RUM is maximally rational.

\bigskip

\noindent \textsc{Remark.}\textit{\textbf{\ }}The converse of Proposition
3.3 is false, because an almost moderately s-transitive stochastic choice
function may still exhibit inconsist choices across nested menus, thereby
violating either $\lambda $-Stochastic Chernoff or $\lambda $-Stochastic
Condorcet Axioms for some $\lambda .$ However, as we will see in Section
4.1, if we restrict our attention to pairwise choice sets (as it is often
done in within-subject choice experiments), then every moderately
s-transitive stochastic choice function is, perforce, maximally rational.

\bigskip

\noindent \textsc{Remark.}\textit{\textbf{\ }}A maximally rational
stochastic choice function need not be strongly s-transitive. This is hardly
a cause for concern. As discussed at length by He and Natenzon (2023),
strong s-transitivity is a very demanding condition. Not only that it is not
satisfied by most random choice models, but it is empirically refuted by
numerous experiments. There are also rather intuitive illustrations of why
\textquotedblleft rational\textquotedblright\ behavior may not abide by this
condition. We will consider one such illustration in Example 3.5 below.

\subsection{Supplementary Comments}

\noindent \textsf{\textbf{Comparison with the Apesteguia-Ballester Swap Index%
}}. To the best of our knowledge, the only measure of rationality in the
literature that applies to stochastic choice functions is the \textit{swap
index} of Apesteguia and Ballester (2015). In the present framework, we can
define this index as the map $I_{\text{swap}}:$ \textsf{\textbf{scf}}$%
(X)\rightarrow \lbrack 0,\infty )$ with 
\begin{equation*}
I_{\text{swap}}(\mathbb{P}):=\min \sum_{S\in \mathfrak{X}}\sum_{x\in S}%
\mathbb{P}(x,S)\left\vert \{w\in S:w\succ x\}\right\vert
\end{equation*}%
where the minimum is taken over all linear orders $\succcurlyeq $ on $X.$ ($%
I_{\text{swap}}$ is an index of \textit{ir}rationality; $I_{\text{swap}}(%
\mathbb{P})>I_{\text{swap}}(\mathbb{Q})$ entails that $\mathbb{P}$ is less
rational than $\mathbb{Q}$.)

Apesteguia and Ballester (2015) provide an axiomatic characterization of the
swap index $I_{\text{swap}}$, and present nice examples that demonstrate
that the linear order $\succcurlyeq $ that achieves the minimum in the
definition of $I_{\text{swap}}$ may well be considered as a welfare
criterion for the subject individual. However, this index, as well as the
general approach followed in that paper, ignores issues like indifference
and indecisiveness that may well be the real cause behind the random choice
behavior of a person. Perhaps the most straightforward illustration of this
point is provided by the following example.

\bigskip

\noindent \textsc{Example 3.3}. Let $X:=\{x,y\}$ and consider an individual
who is indifferent between (or indecisive about) the options $x$ and $y.$
This is a standard scenario, and of course, the individual in question is
surely rational in the deterministic sense. Now it is only natural that this
individual finalizes her choice between $x$ and $y$ by means of a fair coin
toss, which means her stochastic choice function $\mathbb{P}$ on $\mathfrak{X%
}$ is of the form $\mathbb{P}(x,\{x,y\})=\frac{1}{2}=\mathbb{P}(y,\{x,y\})$.
Trivially, this $\mathbb{P}$ is maximally rational (according to $%
\trianglerighteq _{\text{rat}}$). And yet, $I_{\text{swap}}(\mathbb{P})=%
\frac{1}{2}$ here, which is the largest value in $I_{\text{swap}}($\textsf{%
\textbf{scf}}$(X)$). Thus, $I_{\text{swap}}$ declares $\mathbb{P}$ the 
\textit{least} rational random choice behavior in the context of this
example. $\square $

\bigskip

Another point that distinguishes our approach from that of Apesteguia and
Ballester (2015) is that while our rationality ordering/index is geared
toward detecting violations of stochastic transitivity, the swap index
largely ignores this issue. Indeed, the swap index may well declare a
strongly s-transitive stochastic choice function less rational than a
stochastic choice function that is not even weakly s-transitive. The
following example provides an illustration.

\bigskip

\noindent \textsc{Example 3.4}. Let $X:=\{x,y,z\}$ and define $%
u:X\rightarrow \mathbb{R}$ by $u(x):=20,$ $u(y):=19,$ and $u(z):=18.$ Let $%
\mathbb{P}\in $ \textsf{\textbf{scf}}$(X)$ be the Luce model induced by $u,$
that is, $\mathbb{P}(a,S):=u(a)/\sum\nolimits_{b\in S}u(b)$ for any $a\in S$%
. Then, $\mathbb{P}(x,\{x,y\})=\frac{20}{39},$ $\mathbb{P}(y,\{y,z\})=\frac{%
19}{37}$ and $\mathbb{P}(z,\{x,z\})=\frac{18}{38}$. As is well-known, the
Luce model is strongly s-transitive. Now take any stochastic choice function 
$\mathbb{Q}$ on $\mathfrak{X}$ with $\mathbb{Q}(x,\{x,y\})=\mathbb{Q}%
(y,\{y,z\})=\mathbb{Q}(z,\{x,z\})=0.7.$ Then, $\mathbb{Q}$ is not even
weakly s-transitive, but the swap index declares $\mathbb{Q}$
\textquotedblleft more rational\textquotedblright\ than $\mathbb{P}$. By
contrast, we have $\mathbb{P}\vartriangleright _{\text{rat}}\mathbb{Q}.$ In
fact, as we shall see in Section 5.2, every Luce model is maximally
rational. $\square $

\bigskip

These examples suggest that the rationality ordering $\trianglerighteq _{%
\text{rat}}$, as well as the index $I_{\text{rat}}$, are after measuring
different aspects of rationality than the swap index. We aim at
understanding the extent to which the choice probabilities of an individual
are \textit{consistently} generated (in the sense that any deterministic
reflection of this behavior appears rational). As Theorem 2.3 demonstrates,
this entails that violations of stochastic transitivity and/or
inconsistencies of choice probabilities across nested menus lower the
rationality ranking of the given stochastic choice function. In contrast, $%
I_{\text{swap}}$ examines the extent to which these probabilities put us
away from a deterministic choice behavior that is rational in the standard
sense (but without permitting indifferences or incomparabilities). As such,
it appears that the Apesteguia-Ballester approach is best suited for
investigating the rationality of deterministic choice procedures. After all,
a necessary condition for a stochastic choice function $\mathbb{P}$ to be
rendered maximally rational by $I_{\text{swap}}$ is that $\mathbb{P}$ be $%
\{0,1\}$-valued. Put differently, $I_{\text{swap}}$ never views a random
choice rule that assigns positive probabilities to two distinct items in a
menu, as fully rational.\footnote{%
It may be interesting to study a variation of the swap index where the
minimum is taken over all preorders on $X$ (as opposed to linear orders). On
the upside, this would allow for indifference and incomparabilities of the
alternatives, and make the Apesteguia-Ballester approach better apply to
stochastic choice rules. On the downside, it would entail severe
computational difficulties. The swap index is already difficult to compute;
when $X$ has $k$ elements, the minimum in the definition of $I_{\text{swap}}$
is taken over $k!$ linear orders. If we allow for preorders, this number
rises to the so-called $k$th ordered Bell number. For $k\geq 15,$ this
number is approximately $\frac{k!}{2(\log 2)^{k+1}}$ which exceeds $k!$ by
an exponential factor.}

\bigskip

\noindent \textsf{\textbf{Variations on }}$\trianglerighteq _{\text{rat}}$.
Our rationality ordering $\trianglerighteq _{\text{rat}}$ is based on two
premises: (1) Assigning to every $\mathbb{P}\in $ \textsf{\textbf{scf}}$(X)$
its associated Fishburn family; and (2) comparing any two $\mathbb{P}$ and $%
\mathbb{Q}$ in \textsf{\textbf{scf}}$(X)$ by comparing the rationality of $%
C_{\mathbb{P}\text{,}\lambda }$ and $C_{\mathbb{Q}\text{,}\lambda }$ for
each $\lambda \in (0,1].$ Altering the specifications used in (1) and (2)
leads to new methods of making rationality comparisons. First, one may
choose to \textquotedblleft approximate\textquotedblright\ a stochastic
choice function $\mathbb{P}$ by using a family distinct from the Fishburn
family. Provided that one adopts reasonable notions of \textquotedblleft
approximation,\textquotedblright\ this may well yield useful measurement
procedures, but we do not pursue this direction in this paper. Second, one
may keep (1) as is, but adopt a more sophisticated method of comparing the
rationality of $C_{\mathbb{P}\text{,}\lambda }$ and $C_{\mathbb{Q}\text{,}%
\lambda }$ for each $\lambda \in (0,1],$ drawing from the literature on
comparative measurement of rationality of deterministic choice
correspondences. To illustrate, let us use the Houtman-Maks index for this
purpose. This index, which we denote by $I_{\text{H-M}}$, assigns to any
choice correspondence $C$ on $\mathfrak{X}$ the minimum number of menus we
need to eliminate in order to ensure that $C$ is rational,\footnote{%
Our definition of \textquotedblleft rational\textquotedblright\ is not the
same as that of Houtman and Maks (1985) who use this term in the sense of
the maximization of a \textit{complete} preference relation. This is,
however, inconsequential for the point we wish to make here.} that is, 
\begin{equation*}
I_{\text{H-M}}(C):=\min \{\left\vert \mathfrak{Y}\right\vert :\mathfrak{%
Y\subseteq X}\text{ and }C|_{\mathfrak{X}\backslash \mathfrak{Y}}\text{ is
rational}\}\text{.}
\end{equation*}%
We may combine this index with our partial ordering approach by defining a
new preorder $\trianglerighteq _{\text{rat}}^{\prime }$ on \textsf{\textbf{%
scf}}$(X)$ with 
\begin{equation*}
\mathbb{P}\trianglerighteq _{\text{rat}}^{\prime }\mathbb{Q}\text{\hspace{%
0.2in}iff}\hspace{0.2in}I_{\text{H-M}}(C_{\mathbb{P},\lambda })\leq I_{\text{%
H-M}}(C_{\mathbb{Q},\lambda })\text{ for every }\lambda \in (0,1]\text{.}
\end{equation*}%
We do not investigate the applications of these types of hybrid orderings in
the present paper. However, we note that this method results in a coarser
method of ordering than the one proposed here. That is, $\trianglerighteq _{%
\text{rat}}$ contains $\trianglerighteq _{\text{rat}}^{\prime }$, but not
conversely.

\bigskip

\noindent \textsf{\textbf{Completeness of Preferences and }}$%
\trianglerighteq _{\text{rat}}$. Recall that we define a choice
correspondence as \textit{rational} in this paper if it can be rationalized
by a possibly incomplete preference relation. As noted earlier, working with
incomplete preferences is particularly relevant in the present framework,
because the probabilistic choice behavior of a rational individual may well come
about due to their indecisiveness about the appeal of various outcomes (Ok
and Tserenjigmid (2022)). To drive this point home, let us say that a choice
correspondence $C$ on $\mathfrak{X}$ is \textit{totally rational} if there
is a complete preference relation $\succsim $ on $X$ such that $C=\max
(\cdot ,\succsim )$, and define the preorder $\trianglerighteq _{\text{rat}%
}^{\prime \prime }$ on \textsf{\textbf{scf}}$(X)$ with $\mathbb{P}%
\trianglerighteq _{\text{rat}}^{\prime \prime }\mathbb{Q}$ iff 
\begin{equation*}
C_{\mathbb{P},\lambda }\text{ is totally rational for every }\lambda \in
(0,1]\text{ for which }C_{\mathbb{Q},\lambda }\text{ is totally rational.}
\end{equation*}%
This preorder is not nested with $\trianglerighteq _{\text{rat}}$, and it
may at first seem like a reasonable alternative to $\trianglerighteq _{\text{%
rat}}$. The following example issues a warning against this contention,
however.

\bigskip

\noindent \textsc{Example 3.5}. We borrow the parable of this example from
Tversky (1972). A person who lives in New York has to choose between a trip
to Paris ($x$) and a trip to Rome ($y$). Suppose they are conflicted between
these options, to the extent that they are ready to settle the problem by
means of a fair coin toss: $\mathbb{P}(x,\{x,y\})=\frac{1}{2}$. Suppose next
this person is offered a new alternative $x_{+}$, which consists of the trip
to Paris with a discount of \$1. It is plain that they prefer $x_{+}$ over $x
$ with certainty: $\mathbb{P}(x_{+},\{x_{+},x\})=1$. However, a discount of
\$1 does not help alleviate the conflict the decision-maker has about the
two trips. It seems only natural that they settle the choice between $x_{+}$
and $y$ exactly as they settled that problem between $x$ and $y,$ that is,
by using a fair coin toss: $\mathbb{P}(x_{+},\{x_{+},y\})=\frac{1}{2}=%
\mathbb{P}(y,\{x_{+},y\})$.\footnote{%
For another parable that justifies this stochastic choice function from the
perspective of collective choice, see Eliaz and Ok (2006).} This random
choice behavior seems quite sensible; there seems to be nothing
\textquotedblleft irrational\textquotedblright\ about it.\ And yet, we have $%
C_{\mathbb{P},\lambda }\{x,y\}=\{x,y\},$ $C_{\mathbb{P},\lambda
}\{x,x_{+}\}=\{x\},$ and $C_{\mathbb{P},\lambda }\{x_{+},y\}=\{x_{+},y\},$
so $C_{\mathbb{P},\lambda }$ is not \emph{totally} rational for any $\lambda
\in (0,1]$. It follows that $\trianglerighteq _{\text{rat}}^{\prime \prime }$
declares $\mathbb{P}$ less rational than \textit{any }stochastic choice
function on $\mathfrak{X}$, an absurd conclusion.\footnote{%
Indeed, we obtain a very similar conclusion even if $\mathbb{P}%
(x_{+},\{x_{+},y\})>\frac{1}{2}$.}

It is worth noting that the conclusion of $\trianglerighteq _{\text{rat}}$
is wildly different from that of $\trianglerighteq _{\text{rat}}^{\prime
\prime }$ in this example. Indeed, $C_{\mathbb{P},\lambda }$ is easily
rationalized by a preference relation $\succsim $ on $\{x,x_{+},y\}$ (that
says $x_{+}\succ x$ but leaves $x$ and $y,$ and $x_{+}$ and $y,$
incomparable) for every $\lambda \in (0,1]$. Thus, $\trianglerighteq _{\text{%
rat}}$ declares $\mathbb{P}$ as maximally rational, which seems to make
perfect sense in this instance.\footnote{%
In this example, $\mathbb{P}$ is not strongly s-transitive. This shows not
only that a maximally rational choice behavior need not be strongly
s-transitive, but that this is for a good reason. He and Natenzon (2023)
reviews other such examples from the literature.}\thinspace $\square $

\bigskip

\noindent \textsf{\textbf{Simplifications for }}$\Lambda (\mathbb{P})$%
\textsf{\textbf{. }}For certain types of models $\mathbb{P}$, one or more of
the sets that make up $\Lambda (\mathbb{P})$ vanish, so the computation of $%
\Lambda (\mathbb{P})$ gets easier. For example, $\lambda $-rationality
analysis simplifies for moderately s-transitive stochastic choice functions
by the following observation:

\bigskip

\noindent \textsc{Lemma 3.4.} ST$(\mathbb{P})=\varnothing $\textit{\ for
every moderately s-transitive }$\mathbb{P}\in $ \textsf{\textbf{scf}}$(X).$

\smallskip 

\noindent \textsc{Proof. }See Appendix.

\bigskip 

This observation reduces the computation of $\Lambda (\mathbb{P})$ to that
of Ch$(\mathbb{P})$ and Con$(\mathbb{P})$ for moderately s-transitive $%
\mathbb{P}$. It applies, for instance, to all Fechnerian models (such as any
RUM with i.i.d. errors and the weak APU model of Fudenberg, Iijima and
Strzalecki (2015)).\footnote{\linespread{1.1} \selectfont$\mathbb{P}\in $ 
\textsf{\textbf{scf}}$(X)$ is said to be \textbf{Fechnerian} if there exist
an increasing $F:\mathbb{R}\rightarrow \lbrack 0,\infty )$ and a (utility)
function $u:X\rightarrow \mathbb{R}$ such that $\mathbb{P}%
(x,\{x,y\})=F(u(x)-u(y))$ for every $x,y\in X$.} Among the non-Fechnerian
models, Lemma 3.4 also applies to many well-known special cases of RUM such
as probit, nested logit, tremble model, and Tversky's elimination-by-aspects
model.

For our next simplification, let us agree to call a $\mathbb{P}\in $ \textsf{%
\textbf{scf}}$(X)$ \textbf{selective with respect to contractions} if for
any $S,T\in \mathfrak{X}$ with $S\subseteq T$ and $x,y\in S,$ 
\begin{equation*}
\mathbb{P}(x,T)>\mathbb{P}(y,T)\hspace{0.2in}\text{implies\hspace{0.2in}}%
\frac{\mathbb{P}(y,T)}{\mathbb{P}(x,T)}\geq \frac{\mathbb{P}(y,S)}{\mathbb{P}%
(x,S)}\text{,}
\end{equation*}%
and \textbf{selective with respect to expansions} if for any $S,T\in 
\mathfrak{X}$ with $S\subseteq T$ and $x,y\in S,$ 
\begin{equation*}
\mathbb{P}(x,S)>\mathbb{P}(y,S)\hspace{0.2in}\text{implies\hspace{0.2in}}%
\frac{\mathbb{P}(y,S)}{\mathbb{P}(x,S)}\geq \frac{\mathbb{P}(y,T)}{\mathbb{P}%
(x,T)}\text{.}
\end{equation*}%
For example, the Luce model is selective with respect to both contractions
and expansions. Moreover, any tremble model of choice is selective with
respect to expansions, while many weak APU models -- for instance, the ones
in which the derivative of the cost function has a log-convex inverse -- are
selective with respect to contractions. The following observations simplify
the $\lambda $-rationality analysis for such models.

\bigskip

\noindent \textsc{Lemma 3.5.} Con$(\mathbb{P})=\varnothing $\textit{\ for
every $\mathbb{P}\in $ \textsf{\textbf{scf}}$(X)$ that is selective with
respect to contractions.}

\bigskip

\noindent \textsc{Lemma 3.6.} Ch$(\mathbb{P})=\varnothing $\textit{\ for
every $\mathbb{P}\in $ \textsf{\textbf{scf}}$(X)$ that is selective with
respect to expansions.}

\bigskip

The proofs of these results are obtained from the definitions routinely, and
are thus omitted.

\bigskip 

\noindent \textbf{On the Meaning of Using Different $\lambda $s.} It is
essential for our rationality order to use all values for $\lambda $ in
(0,1] to capture violations of rationality that arise from inconsistent
choices across nested menus as well as from the failure of stochastic
transitivity. We give two examples to illustrate this. 

Put $X:=\{x,y,z\}.$ Consider the stochastic choice function $\mathbb{P}$ on $%
\mathfrak{X}$ such that $\mathbb{P}(x,X)=1-\varepsilon $, $\mathbb{P}%
(y,X)=\varepsilon $, and $\mathbb{P}(x,\{x,y\})=\mathbb{P}(y,\{y,z\})=%
\mathbb{P}(x,\{x,z\})=1,$ where $\varepsilon $ is a small positive number.
Intuitively, $x$ and $z$ are the best and worst options for the involved
person, respectively, and the second best option $y$ is chosen over $x$ from 
$X$ with probability $\varepsilon $. It is easily found that $\Lambda (%
\mathbb{P})=(0,\frac{\varepsilon }{1-\varepsilon }].$ In this example,
therefore, small values of $\lambda $ capture potential inconsistencies of
choices across menus.

Consider next a person whose stochastic choice function $\mathbb{Q}$ on $%
\mathfrak{X}$ satisfies $\mathbb{Q}(x,X)=\mathbb{Q}(y,X)=\frac{1}{3}$ and $%
\mathbb{Q}(x,\{x,y\})=\mathbb{Q}(y,\{y,z\})=\mathbb{Q}(z,\{x,z\})=\frac{1}{2}%
+\varepsilon $. Intuitively, all three alternatives are equally good, but $x$
(resp., $y$ and $z$) is chosen more frequently than $y$ (resp., than $z$ and 
$x$) with probability $\varepsilon $ so that $\mathbb{Q}$ violates weak
s-transitivity. Then, $\Lambda (\mathbb{Q})=(\frac{.5-\varepsilon }{%
.5+\varepsilon },1].$ In this example, therefore, large values of $\lambda $
capture violations of s-transitivity.

\bigskip 

\noindent \textbf{On the Consistency of Tie-Breaking Rules.} We are of the
opinion that uniform randomization across indifferent alternatives should be
considered rational and our rationality order concludes the same (Example
3.2). On the other hand, a decision maker may finalize their choice among
items that they are indifferent about by using a (random or otherwise)
tie-breaking rule. This may change the structure of their random choice
behavior, and detract from rationality. In particular, the present approach
to measuring stochastic rationality requires tie-breaking rules be
consistent across nested menus (as in the Luce model). If $x$ and $y$ are
indifferent, and $x$ is chosen more frequently than $y$ in large menus, then
maximal rationality demands that $x$ be chosen more frequently than $y$ in
smaller menus as well.

\section{Pairwise Choice Situations}

\subsection{Comparative Rationality with Binary Choice Data}

We have so far worked within the standard framework of stochastic choice
theory which presumes that the analyst is given the choice
probabilities/frequencies of an individual at every feasible menu. This
framework is suitable for making rationality comparisons between theoretical
models of random choice, but it is too demanding when it comes to
experimental applications in which one has only limited data sets. After
all, even with as few as 10 alternatives, it is basically impossible to
collect choice data over all possible menus. In fact, other than very few
exceptions, all within-subject repeated/random choice experiments work with
a collection of menus that contain only two options (i.e., binary menus).%
\footnote{%
See, among many others, Hey and Orme (1994), Hey (2001), Agranov and
Ortoleva (2017), and Feldman and Rehbeck (2022).} Fortunately, the
methodology we introduced in Section 3 not only readily adapts to this case,
but it actually becomes easier to apply.

\bigskip 

\noindent \textbf{Choice with a Collection of Binary Menus. }Let $\mathfrak{X%
}_{2}$ stand for the collection of all binary menus of $X$, and $\mathfrak{Y}
$ an arbitrary nonempty subset of $\mathfrak{X}_{2}$. By a \textbf{choice
correspondence} on $\mathfrak{Y},$ we mean a set-valued map $C\!:\mathfrak{Y}%
\rightarrow \mathfrak{X}$ such that $C(S)\subseteq S$ for each $S\in 
\mathfrak{Y}$. As soon as we replace $\mathfrak{X}$ with $\mathfrak{Y}$, the
notion of being rational applies to such correspondences \textit{verbatim},
while the Chernoff and Condorcet axioms are trivially satisfied. However, in
the context of this sort of a restricted domain, we need to modify the
No-Cycle Axiom as follows:

\bigskip 

\noindent \textbf{Acyclicity.\ }For every $x_{1},\ldots ,x_{n}\in X$ such
that $\{x_{i},x_{i+1(\text{mod n})}\}\in \mathfrak{Y}$ for all $i=1,...,n,$ 
\begin{equation*}
C(\{x_{1},x_{2}\})=\{x_{1}\},...,C(\{x_{n-1},x_{n}\})=\{x_{n-1}\}\text{%
\hspace{0.2in}imply}\hspace{0.2in}C(\{x_{1},x_{n}\})=\{x_{1}\}.
\end{equation*}

\smallskip 

It is an easy exercise to show that this property characterizes rationality;
we omit the proof.

\bigskip 

\noindent \textsc{Proposition 4.1.} \textit{A choice correspondence on }$%
\mathfrak{Y}$ \textit{is rational if, and only if, it satisfies Acyclicity.} 

\bigskip 

By a \textbf{stochastic choice function} on $\mathfrak{Y},$ we mean a
function $\mathbb{P}:X\times \mathfrak{Y}\rightarrow \lbrack 0,1]$ such that 
$\mathbb{P}(x,\{x,y\})+\mathbb{P}(y,\{x,y\})=1$ and $\mathbb{P}(z,\{x,y\})=0$
for every $\{x,y\}\in \mathfrak{Y}$ and $z\in X\backslash \{x,y\}.$ The
collection of all such functions is denoted by \textsf{\textbf{scf}}$_{%
\mathfrak{Y}}(X).$ In turn, for any $\mathbb{P}\in $ \textsf{\textbf{scf}}$_{%
\mathfrak{Y}}(X)$ and $\lambda \in (0,1],$ we say that $\mathbb{P}$ is 
\textbf{$\lambda $-rational} if the choice correspondence $C_{\mathbb{P}%
,\lambda }$ on $\mathfrak{Y}$ is rational. In view of Proposition 4.1,
characterization of this requires modifying the $\lambda $-Stochastic
Transitivity Axiom as follows:

\bigskip 

\noindent $\mathcal{\lambda }$\textbf{-Acyclicity.\ }For every $x_{1},\ldots
,x_{n}\in X$ such that $\{x_{i},x_{i+1(\text{mod n})}\}\in \mathfrak{Y}$ for
all $i=1,...,n,$ 
\begin{equation*}
\mathbb{P}^{\ast }(x_{2},\{x_{1},x_{2}\})<\lambda ,\text{ }...,\text{ }%
\mathbb{P}^{\ast }(x_{n},\{x_{n-1},x_{n}\})<\lambda 
\end{equation*}%
imply $\mathbb{P}^{\ast }(x_{n},\{x_{1},x_{n}\})<\lambda .$

\bigskip 

We then have the following modification of Theorem 2.3, which follows from
the definitions in the obvious way.

\bigskip 

\noindent \textsc{Proposition 4.2.} \textit{For any }$\lambda \in (0,1],$%
\textit{\ a stochastic choice function} \textit{on} $\mathfrak{Y}$\textit{\
is }$\lambda $\textit{-rational if, and only if, it satisfies }$\lambda $%
\textit{-Acyclicity.}

\bigskip 

Next, we define the \textbf{comparative rationality ordering}\textbf{\ }$%
\trianglerighteq _{\text{rat}}$ on \textsf{\textbf{scf}}$_{\mathfrak{Y}}(X)$
by (\ref{ratt}). Besides, Proposition 4.2 prompts defining the \textbf{%
acyclicity set} of $\mathbb{P}$ as%
\begin{equation}
\text{AC}(\mathbb{P}):=\bigcup \left( \max_{i=1,...,n-1}\mathbb{P}^{\ast
}(x_{i+1},\{x_{i},x_{i+1}\},\mathbb{P}^{\ast }(x_{n},\{x_{1},x_{n}\})\right]
,  \label{six}
\end{equation}%
where the union is taken over all integers $n\geq 2$ and $x_{1},\ldots
,x_{n}\in X$ such that $\{x_{i},x_{i+1(\text{mod n})}\}\in \mathfrak{Y}$ for
each $i=1,...,n.$ This leads to the following characterization:%
\begin{equation*}
\mathbb{P}\trianglerighteq _{\text{rat}}\mathbb{Q}\hspace{0.2in}\text{iff%
\hspace{0.2in}AC}(\mathbb{P})\subseteq \text{AC}(\mathbb{Q})
\end{equation*}%
for every $\mathbb{P},\mathbb{Q}\in $ \textsf{\textbf{scf}}$_{\mathfrak{Y}%
}(X)$. In turn, our rationality index takes the form%
\begin{equation*}
I_{\text{rat}}(\mathbb{P}):=1-\text{Leb}(\text{AC}(\mathbb{P}))
\end{equation*}%
for every $\mathbb{P}\in $ \textsf{\textbf{scf}}$_{\mathfrak{Y}}(X)$. 

These observations render our measurement approach readily applicable to
experimental data sets in which subjects make choices from several binary
menus, may these be random or repeated. A demonstration of this will be
presented in Section 5.4.

\bigskip 

\noindent \textbf{Choice with All Binary Menus. }When the choice data
pertains to all pairwise choice situations, that is, $\mathfrak{Y}=\mathfrak{%
X}_{2}$, the exercise at hand is thoroughly simplified. Proposition 4.1 in
that case takes the following form: \textit{A choice correspondence on }$%
\mathfrak{X}_{2}$ \textit{is rational iff it satisfies the No-Cycle Axiom. }%
Similarly, Proposition 4.2 becomes: \textit{For any }$\lambda \in (0,1],$%
\textit{\ a stochastic choice function} \textit{on} $\mathfrak{X}_{2}$ 
\textit{\ is }$\lambda $\textit{-rational iff it satisfies the }$\lambda $%
\textit{-Stochastic Transitivity Axiom. }In other words, when $\mathfrak{Y}=%
\mathfrak{X}_{2}$, the Acyclicity and Stochastic Transitivity sets coincide,
and hence 
\begin{equation*}
\mathbb{P}\trianglerighteq _{\text{rat}}\mathbb{Q}\hspace{0.2in}\text{iff%
\hspace{0.2in}ST}(\mathbb{P})\subseteq \text{ST}(\mathbb{Q}),
\end{equation*}%
and $I_{\text{rat}}(\mathbb{P})=1-$ Leb$($ST$(\mathbb{P})),$ for every
stochastic choice functions $\mathbb{P}$ and $\mathbb{Q}$ on $\mathfrak{X}%
_{2}$. As ST$(\mathbb{P})=\varnothing $ for every moderately s-transitive $%
\mathbb{P}\in $ \textsf{\textbf{scf}}$_{\mathfrak{X}_{2}}(X),$ therefore,
such functions are automatically maximally rational. Moreover, the reasoning
behind Proposition 3.3 applies to this restricted domain without alteration.
Thus: 
\begin{equation*}
\text{moderate s-transtivity}\Longrightarrow \text{maximal rationality}%
\Longrightarrow \text{almost moderate s-transitivity}
\end{equation*}%
for every $\mathbb{P}\in $ \textsf{\textbf{scf}}$_{\mathfrak{X}_{2}}(X)$.
This shows that the difference between maximal rationality and moderate
s-transitivity is really minor in the case of pairwise choice situations.
The following is an immediate application of this observation.

\bigskip

\noindent \textsc{Example 4.1}. The \textbf{moderate utility model} (MUM) of
He and Natenzon (2023) is any stochastic choice function $\mathbb{P}$ on $%
\mathfrak{X}_{2}$ of the form 
\begin{equation*}
\mathbb{P}(x,\{x,y\})=F\left( \frac{u(x)-u(y)}{d(x,y)}\right) ,\hspace{0.1in}%
\hspace{0.1in}\hspace{0.1in}x,y\in X,
\end{equation*}%
where $F:\mathbb{R}\rightarrow \lbrack 0,\infty )$ is a strictly increasing
function with $F(t)=1-F(-t)$ for every $t\in \mathbb{R},$ $u:X\rightarrow 
\mathbb{R}$ is a (utility) function, and $d$ is any metric on $X.$ It is not
difficult to show that any MUM is moderately s-stochastic. (He and Natenzon
(2023) in fact prove that the converse of this is also true, that is, every
moderately s-stochastic $\mathbb{P}\in $ \textsf{\textbf{scf}}$_{2}(X)$ is a
MUM.) In view of the discussion above, therefore, we conclude: \textit{Every
MUM is maximally rational.}

Consequently, all Fechnerian models (e.g., RUM with i.i.d. errors and the
weak APU model) and many well-known RUMs (e.g., probit, nested logit model,
tremble model, and Tversky's elimination-by-aspects model) are maximally
rational on $\mathfrak{X}_{2}$ as they are special instances of the MUM. $%
\square $

\subsection{Computation of Acyclicity Sets}

As the number of potential cycles increases exponentially in the number of
alternatives, one may worry that it may not be possible to compute AC$(%
\mathbb{P})$ for a stochastic choice function $\mathbb{P}$ on some $%
\mathfrak{Y}$ efficiently. Insofar as choice experiments in the lab are
concerned, this is not a cause for concern. As they have to record the
choices of the subjects in a limited amount of time, such experiments
invariably keep the number of binary menus, that is, $\left\vert \mathfrak{Y}%
\right\vert $, small, often less than 25. This, in turn, forces severe
bounds on the numbers of alternatives and potential cycles.\footnote{%
When $|X|\leq 10$, there are no computational issues as there are at most $%
\frac{10!}{2}$ potential cycles. However, even if $|X|>10$, as long as $|%
\mathfrak{Y}|$ remains below $30$, the computation of the Acyclicity sets is
not difficult since in that case there are at most 100 million potential
cycles. As a matter of fact, in experimental practice, the number of
potential cycles is significantly less. For example, in Hey and Orme (1994),
there are only 50 potential cycles since there are 25 menus and each
alternative only appears twice.} Unlike the exercise of measuring deviations
from RUM, even brute-force calculations of Acyclicity sets are thus
computationally feasible with experimental data sets.

We next present an algorithm for calculating AC$(\mathbb{P})$ for any $%
\mathbb{P}\in $ \textsf{\textbf{scf}}$_{\mathfrak{Y}}(X)$ which is
surprisingly fast even when $X$ and $\mathfrak{Y}$ are reasonably large. In
the next section we will use this algorithm in the context of the
experiments of Feldman and Rehbeck (2022) who provide the largest random
choice data set we are aware of.

\bigskip 

\noindent \textbf{An Algorithm for Computing }AC$(\mathbb{P})$\textbf{.} In
repeated choice experiments, empirical choice probabilities are, perforce,
rational numbers. In random choice experiments, on the other hand, subjects
are given a set of probabilities to assign to each alternative in a menu,
and naturally, these potential probabilities are chosen (by the
experimenter) as rational numbers between $0$ and $1.$ Regardless of the
exact nature of the choice experiment at hand, therefore, observed choice
probabilities, and hence the values of $\mathbb{P}^{\ast }$, of any subject
come from a finite set of rational numbers in $[0,1]$. Let us define the set
of all observed $\mathbb{P}^{\ast }$ values by $\Theta ,$ that is, 
\begin{equation*}
\Theta :=\{\mathbb{P}^{\ast }(x,\{x,y\}):\{x,y\}\in \mathfrak{Y}\mathcal{\}}%
\text{.}
\end{equation*}%
The experimenter is able to control the size of this set by design.

The initial stages of our algorithm are parametric over $\Theta.$ For each
(fixed) $\theta \in \Theta ,$ it proceeds as follows:

[Step 1.$\theta $] Let $G_{\theta }$ be the digraph whose vertex set is $X$
and in which there is a directed edge between the vertices $x$ and $y$ iff $%
\mathbb{P}^{\ast }(a,\{a,b\})\leq \theta $. For any $x\in X,$ we denote by $%
R_{\theta }(x)$ the set of all $y\in X$ that is reachable from $x$ (which
means there is a directed path from $x$ to $y$ in $G_{\theta }$). This first
step of our algorithm is to compute each reachability set $R_{\theta }(x)$.
There are various ways of doing this efficiently. For example, the
depth-first search algorithm is commonly used for this purpose.

[Step 2.$\theta $] Take any $x,y\in X$ with $\{x,y\}\in \mathfrak{Y}$ and $%
\mathbb{P}^{\ast }(x,\{x,y\})>\theta .$ (If there is no such $x$ and $y,$ we
set $J_{\theta }=\varnothing ,$ which means we stop this step with no
output.) Now for every $a,b\in X$ with $\{a,b\}\in \mathfrak{Y}$ and $%
\mathbb{P}^{\ast }(a,\{a,b\})=\theta ,$ we check if $a\in R_{\theta }(x)$
and $y\in R_{\theta }(b).$ If there is no such $a$ and $b,$ we set $%
J_{\theta }(x,y)=\varnothing $. Otherwise, we set $J_{\theta }(x,y)=(\theta ,%
\mathbb{P}^{\ast }(x,\{x,y\})]$ which must be contained within AC$(\mathbb{P}%
)$. We conclude this step by collecting all the intervals obtained this way,
that is, recording%
\begin{equation*}
J_{\theta }=\bigcup \{J_{\theta }(x,y):\{x,y\}\in \mathfrak{Y}\text{ and }%
\mathbb{P}^{\ast }(x,\{x,y\})>\theta \}\text{.}
\end{equation*}

[Step 3] In this final stage of the algorithm, we simply adjoin the
intervals found in Step 2.$\theta $ across all $\theta \in \Theta $, and
obtain the Acyclicity set we are after exactly:%
\begin{equation*}
\text{AC}(\mathbb{P})=\bigcup_{\theta \in \Theta }J_{\theta }\text{.}
\end{equation*}

Despite its brute-force elements, this algorithm can be executed in
polynomial time. Indeed, its time complexity is only $O(\left\vert \Theta
\right\vert \times \left\vert \mathfrak{Y}\right\vert ^{2})$.

In Section 5.4.2 we will apply this algorithm in the context of the
experimental data provided by Feldman and Rehbeck (2022). In that
experiment, the number of alternatives is $27$ (i.e. $\left\vert
X\right\vert =27$), and the number of binary menus is $79$ (i.e., $%
\left\vert \mathfrak{Y}\right\vert =79$). Besides, $\Theta $ is a subset of
the set of all fractions $\frac{k-1}{101-k}$ where $k=1,...,50,$ so $%
\left\vert \Theta \right\vert \leq 50$. The data provided by Feldman and
Rehbeck constitutes the largest random choice data set to date, but our
algorithm requires very little computing power to deal with it.

\section{Applications}

\subsection{Random Utility Models}

Being the workhorse of discrete choice theory and empirical industrial
organization, the random utility model (RUM) is the most well-known
stochastic choice model in economics. However, in that role, RUM is not an
individual decision theory, for it uses choice data in the aggregate. It
presumes that the society is partitioned in such a way that any two
individuals in any given cell of the partition have the same (deterministic)
utility function, and the relative sizes of these cells are known. In turn,
it records the relative frequencies of the utility-maximizing choices of the
individuals in every given menu. From this viewpoint, therefore, the
\textquotedblleft stochastic\textquotedblright\ nature of the model applies
only to a randomly chosen individual in the society, not to any one
particular person. After all, every individual in the society is taken as a
rational person in the standard, deterministic sense. This is perhaps why
many economists regard the RUM as a \textquotedblleft
rational\textquotedblright\ choice model.\footnote{%
When one considers RUM from this aggregate-choice point of view, asking
questions about its rationality becomes questions about the rationality of
the society as a whole, and as such, they are not of great interest. This
is, of course, not a novel perspective. Buchanan (1954), for instance,
famously argued that because society is not an \textquotedblleft organic
entity,\textquotedblright\ it is meaningless to contemplate about its
(collective) rationality. }

When we consider the RUM as a model of \textit{individual} decision-making,
things are different. It is then not self-evident why one should regard any
such model as corresponding to the behavior of a rational individual. As
discussed in the Introduction, it is easy to come up with convincing stories
(as in the parable of the \textquotedblleft crazy
politician\textquotedblright ) in which the behavior generated by certain
types of RUM does not appear rational at all. Indeed, the axiomatic
foundations of the RUM are categorically different than those of the
deterministic rational choice model. These come in the form of the
Block-Marshak inequalities (Falmagne (1978)) or the Axiom of Revealed
Stochastic Preference (McFadden (2005)), but the behavioral interpretations
of these properties are far from transparent.\footnote{%
This difficulty is well-recognized in the literature and has led many
decision theorists to characterize special cases of the RUM that emanate
from simpler behavioral properties. See, for instance, Gul and Pesendorfer
(2004), and Apestguia, Ballester and Lu (2014).} To add insult to injury, it
is well known that a RUM may fail even weak s-transitivity, the most basic
stochastic transitivity requirement considered in the literature.

These observations are not meant to criticize the RUM. If anything, they
entail that the RUM, as a decision-making model of an individual, may be
flexible enough to capture one's boundedly rational choice behavior as well.
Furthermore, obviously, this discussion does not say that \textit{no }RUM
can be viewed as rational. It only opens up the question of determining
which types of RUMs can be thought of as rational. In what follows, we use
the measurement methodology introduced in Section 3 to look at this issue
rigorously, and try to understand when a random utility model is considered
rational, and when it is not, according to the rationality ordering $%
\trianglerighteq _{\text{rat}}$. We first look at the case of RUMs with two
utility functions, as the results in that case are particularly transparent.
The general case is taken up subsequently.

\subsubsection{The Uniform Dual Random Utility Model}

Let $u$ and $v$ be two injective real maps on $X$ and $\theta :\mathfrak{X}%
\rightarrow \lbrack 0,1]$ any function. The \textbf{dual random utility model%
} (dRUM) induced by $(u,v,\theta )$ is the stochastic choice function $%
\mathbb{P}\in $ \textsf{\textbf{scf}}$(X)$ with%
\begin{equation*}
\mathbb{P}(\cdot ,S)=\theta (S)\,\mathbf{1}_{\arg \max u(S)}+(1-\theta (S))\,%
\mathbf{1}_{\arg \max v(S)}
\end{equation*}%
for every $S\in \mathfrak{X}$. When $\theta $ is a constant function here,
that is, 
\begin{equation*}
\mathbb{P}(x,S)=\theta\, \mathbf{1}_{\arg \max u(S)}(x)+(1-\theta )\,\mathbf{%
1}_{\arg \max v(S)}(x)
\end{equation*}%
for every $x\in X$ and $S\in \mathfrak{X},$ we refer to this model as the 
\textbf{uniform dRUM }induced by\textbf{\ }$(u,v,\theta )$. (This model is
said to be \textit{non-degenerate} if $\theta \in (0,1)$.) Very nice
characterizations of these models have been provided by Manzini and Mariotti
(2018).

Let $\mathbb{P}$ be a uniform dRUM induced by $(u,v,\theta )$, and
relabelling if necessary, let us assume $\theta \geq \frac{1}{2}$. Then $%
\mathbb{P}^{\ast }(x,S)$ may equal either $0,\frac{1-\theta }{\theta },$ or $%
1$, for any $x\in X$ and $S\in \mathfrak{X}$. Now, it is readily checked
that Ch$(\mathbb{P})=\varnothing $. Moreover, if $\theta >\frac{1}{2}$, then 
$\mathbb{P}$ is moderately s-transitive, so ST$(\mathbb{P})=\varnothing $ by
Lemma 3.4. If $\theta =\frac{1}{2}$, then $\mathbb{P}^{\ast }$ is $\{0,1\}$%
-valued, so for any $x,y,z\in X,$ the interval $(\max \{\mathbb{P}^{\ast
}(y,\{x,y\}),\mathbb{P}^{\ast }(z,\{y,z\})\},$ $\mathbb{P}^{\ast
}(z,\{x,z\})]$ is nonempty only if both $\mathbb{P}^{\ast }(y,\{x,y\})$ and $%
\mathbb{P}^{\ast }(z,\{y,z\})$ are zero, but the latter means $%
u(x)>u(y)>u(z) $ and $v(x)>v(y)>v(z),$ whence $\mathbb{P}^{\ast
}(z,\{x,z\})=0,$ so this interval must be empty. Thus, we have ST$(\mathbb{P}%
)=\varnothing $ when $\theta =\frac{1}{2}$ as well.

It follows that, for uniform dRUMs, rationality comparisons on the basis of $%
\trianglerighteq _{\text{rat}}$ depend only on the associated Condorcet
sets. To understand the structure of these, we introduce the following
concept: The real maps $u$ and $v$ on $X$ are said to be \textbf{consistent
over triplets} if there do not exist three alternatives $x,$ $y$ and $z$ in $%
X$ such that 
\begin{equation}
u(x)>u(y)>u(z)\hspace{0.2in}\text{and\hspace{0.2in}}v(z)>v(y)>v(x).
\label{con}
\end{equation}

Then we have the following result:

\bigskip

\noindent \textsc{Proposition 5.1.} \textit{Let $\mathbb{P}$ be the uniform
dRUM on} $\mathfrak{X}$\textit{\ induced by $(u,v,\theta )$ with $\theta
\geq \frac{1}{2}$. If $u$ and $v$ are consistent over triplets, then $%
\mathbb{P}$ is maximally rational. If $u$ and $v$ are not consistent over
triplets, then $\Lambda (\mathbb{P})=(0,\frac{1-\theta }{\theta }].$}

\smallskip 

\noindent \textsc{Proof. }See Appendix.

\bigskip

Quite a bit follows from this finding. Let $\mathbb{P}$ and $\mathbb{Q}$ be
two uniform dRUMs on $\mathfrak{X},$ induced by $(u,v,\alpha )$ and $%
(f,g,\beta ),$ respectively. If $u$ and $v$ are consistent over triplets,
then $\mathbb{P}$ $\trianglerighteq _{\text{rat}}\mathbb{Q}$, and if $f$ and 
$g$ are consistent over triplets, then $\mathbb{Q}$ $\trianglerighteq _{%
\text{rat}}\mathbb{P}$. On the other hand, if neither $u$ and $v,$ nor $f$
and $g,$ are consistent over triplets, then $\max \{\alpha ,1-\alpha \}\geq
\max \{\beta ,1-\beta \}$ implies $\mathbb{P}$ $\trianglerighteq _{\text{rat}%
}\mathbb{Q}$, while the reverse inequality entails $\mathbb{Q}$ $%
\trianglerighteq _{\text{rat}}\mathbb{P}$. In particular, we find that the
rationality ordering $\trianglerighteq _{\text{rat}}$ is complete over the
class of all uniform dRUMs on $\mathfrak{X}.$ Moreover, any uniform dRUM $%
\mathbb{P}$ that is induced by $(u,v,\theta )$ is maximally rational
provided that either $\theta \in \{0,1\}$ or $u$ and $v$ are consistent over
triplets. At the other extreme, $\mathbb{P}$ is minimally rational if $u$
and $v$ are not consistent over triplets and $\theta =\frac{1}{2}$. In
summary:

\bigskip

\noindent \textsc{Corollary 5.2.} \textit{a. Any two uniform dRUMs on} $%
\mathfrak{X}$\textit{\ are $\trianglerighteq _{\text{rat}}$-comparable. }

\textit{b. Let $\mathbb{P}$ and $\mathbb{Q}$ be the uniform dRUMs induced $%
(u,v,\theta)$ and $(u,v,\theta^{\prime })$, respectively. If $\theta\ge
\theta^{\prime }$, then $\mathbb{P}$ $\trianglerighteq _{\text{rat}}\mathbb{Q%
}$.}

\textit{c. Any uniform dRUM induced by utility functions that are consistent
over triplets is maximally rational. }

\textit{d. If a non-degenerate uniform dRUM is induced by utility functions
that are not consistent over triplets, then it is not maximally rational. }

\textit{e. Any uniform dRUM induced $(u,v,\frac{1}{2})$ where $u$ and $v$
are not consistent over triplets, is minimally rational.}

\bigskip

The last two parts of Corollary 5.2 go against the conventional wisdom that
every random utility model is \textquotedblleft rational.\textquotedblright\
Corollary 5.2.d says that a uniform dRUM may be strictly less rational than
some other stochastic choice function. In fact, the dRUM induced by $%
(u,v,\theta ),$ where $u$ and $v$ are not consistent over triplets, and $%
\theta >\frac{1}{2},$ gets less and less rational as $\theta $ decreases to $%
\frac{1}{2}$. The minimal rationality is attained at $\theta =\frac{1}{2}$.

The upshot here is that certain types of random utility models may fail to
be rational in a very strong sense. Unless it induces utility functions
satisfy a suitable consistency condition, such a model may turn out to be
less rational than any other stochastic choice model (according to $%
\trianglerighteq _{\text{rat}}$). Insofar as one wishes to model the
behavior of a rational decision-maker stochastically by means of a random
utility model, they should thus consider imposing such a consistency
condition on the set of utility functions.

\subsubsection{The Random Utility Model}

Take any integer $n\geq 2,$ and put $N:=\{1,...,n\}.$ For each $i\in N,$ let 
$u_{i}$ be an injective real map on $X$, and $\theta _{i}$ a positive number
such that $\theta _{1}+\cdot \cdot \cdot +\theta _{n}=1.$ The \textbf{random
utility model} (RUM) induced by $(u_{i},\theta _{i})_{i\in N}$ is the
stochastic choice function $\mathbb{P}\in $ \textsf{\textbf{scf}}$(X)$ with 
\begin{equation*}
\mathbb{P}(x,S)=\sum_{i\in N}\theta _{i}\mathbf{1}_{\arg \max u_{i}(S)}(x)
\end{equation*}%
for every $x\in X$ and $S\in \mathfrak{X}$. Relabelling if necessary, we
order the $\theta _{i}$s in the decreasing fashion, that is, assume in what
follows that $1>\theta _{1}\geq \cdot \cdot \cdot \geq \theta _{n}>0.$

To generalize the results of the previous subsection, we need to extend the
notion of \textquotedblleft consistency over triplets\textquotedblright\ to
the context of more than two utility functions. We shall use two properties
for this purpose. First, we say that $u_{1},...,u_{n}$ are \textbf{%
consistent over }$(n+1)$\textbf{-tuples} if there do not exist $n+1$
alternatives $x,x_{1},...,x_{n}\in X$ such that 
\begin{equation}
u_{j}(x_{j})>u_{j}(x)>\max_{i\in N\backslash \{j\}}u_{j}(x_{i})\hspace{0.2in}%
\text{for each }j=1,...,n.  \label{conn}
\end{equation}%
Second, we say that $u_{1}$ is \textbf{consistent with }$u_{2},...,u_{n}$%
\textbf{\ over triplets} if there do not exist three alternatives $x,$ $y,$
and $z$ in $X$ such that 
\begin{equation}
u_{1}(x)>u_{1}(y)>u_{1}(z)\hspace{0.2in}\text{and\hspace{0.2in}}%
u_{i}(z)>u_{i}(y)>u_{i}(x)\text{ for each }i=2,...,n.  \label{connn}
\end{equation}%
We note that both of these concepts become identical to \textquotedblleft
consistency over triplets\textquotedblright\ when $n=2.$

The following result generalizes the second part of Proposition 5.1.

\bigskip

\noindent \textsc{Proposition 5.3.} \textit{Let $\mathbb{P}$ be the RUM on} $%
\mathfrak{X}$\textit{\ induced by $(u_{i},\theta _{i})_{i\in N}$ with $%
1>\theta _{1}\geq \cdot \cdot \cdot \geq \theta _{n}>0$. }

\textit{a. If $\frac{1}{2}\geq \theta _{1}$ and $u_{1},...,u_{n}$ are not
consistent over $(n+1)$-tuples, then $\mathbb{P}$ is minimally rational; }

\textit{b. If $\theta _{1}>\frac{1}{2}$ and $u_{1}$ is not consistent with $%
u_{2},...,u_{n}$ over triplets, then $\Lambda (\mathbb{P})=(0,\frac{1-\theta
_{1}}{\theta _{1}}].$}

\bigskip

The following example illustrates that the two consistency conditions used
in Proposition 5.3 are not sufficient for maximal rationality.

\bigskip

\noindent \textsc{Example 5.1.} Let $X:=\{x,y,z\}$, take any maps $u_{1},$ $%
u_{2}$ and $u_{3}$ on $X$ with $u_{1}(x)>u_{1}(y)>u_{1}(z)$, $%
u_{2}(y)>u_{2}(x)>u_{2}(z)$ and $u_{3}(z)>u_{3}(x)>u_{3}(y)$, and set $%
(\theta _{1},\theta _{2},\theta _{3}):=(\frac{2}{3},\frac{1}{6},\frac{1}{6}).
$ If $\mathbb{P}$ is the RUM on $\mathfrak{X}$ induced by $(u_{i},\theta
_{i})_{i\in \{1,2,3\}},$ then $\Lambda (\mathbb{P})=(\frac{1}{5},\frac{1}{4}%
].$ It follows from Proposition 3.2 that $\mathbb{P}$ is not maximally
rational. $\square $

\bigskip

Our next result provides a sufficient condition for a RUM to be maximally
rational.

\bigskip

\noindent \textsc{Proposition 5.4.} \textit{Let $\mathbb{P}$ be the RUM on} $%
\mathfrak{X}$\textit{\ induced by $(u_{i},\theta _{i})_{i\in N}$ with $%
1>\theta _{1}>\frac{1}{2}$. If}%
\begin{equation*}
u_{1}(x)>u_{1}(y)>u_{1}(z)\hspace{0.2in}\text{\textit{implies}\hspace{0.2in}}%
u_{i}(x)>u_{i}(z)\text{ \textit{for all} }i\in N
\end{equation*}%
\textit{for every }$x,y,z\in X$\textit{, then $\mathbb{P}$ is maximally
rational.}

\smallskip 

\noindent \textsc{Proof. }See Appendix.

\bigskip 

Obtaining a complete characterization of all maximally rational RUMs with
three or more utility functions remains an open problem.

\subsection{Luce-Type Models}

Throughout this section, unless stated otherwise, $u:X\rightarrow (0,\infty
) $ is an arbitrary function and $\Gamma :\mathfrak{X}\rightarrow \mathfrak{X%
}$ is any function with $\Gamma (S)\subseteq S$ for every $S\in \mathfrak{X}$%
.

\subsubsection{General Luce Model}

The \textbf{general} \textbf{Luce model} induced by $(u,\Gamma )$ is the
stochastic choice function $\mathbb{P}$ on $\mathfrak{X}$ such that%
\begin{equation*}
\mathbb{P}(x,S):=\frac{u(x)}{\sum\limits_{\omega \in \Gamma (S)}u(\omega )}%
\hspace{0.2in}\text{for any }x\in \Gamma (S),
\end{equation*}%
and $\mathbb{P}(x,S):=0$ for any $x\in X\backslash \Gamma (S)$. We refer to $%
\Gamma $ as the \textbf{constraint correspondence} of the model.

This model was introduced by Ahumada and Ulku (2018) and Echenique and Saito
(2019). We may think of $\Gamma (S)$ as the set of alternatives that survive
a preliminary stage of elimination, or as the set of all alternatives in a
feasible menu $S$ to which the agent somehow limits her attention. The first
interpretation, favored by Echenique and Saito (2019), is in line with
two-stage choice procedures (as in Manzini and Mariotti (2009)), whereas the
second is more in line with the \textit{revealed attention} model of
Masatlioglu, Nakajima, and Ozbay (2012).

A natural question is if among two general Luce models with the same utility
function $u,$ the one with the more restrictive constraint correspondence is
necessarily less rational than the other one. The following two examples
show jointly that this may well fail to be the case.

\bigskip

\noindent \textsc{Example 5.2.} Let $f:X\rightarrow \mathbb{R}$ and define $%
\Gamma :\mathfrak{X}\rightarrow \mathfrak{X}$ by $\Gamma (S):=\arg \max f(S).
$ Then, any general Luce model with the constraint correspondence $\Gamma $
is maximally rational. $\square $

\bigskip

\noindent \textsc{Example 5.3.} Let $X:=\{x,y,z\},$ $u(x)>u(y)>u(z),$ $%
\Gamma \{x,y\}=\{y\}=\Gamma \{y,z\},$ and $\Gamma (X):=\{x,z\}.$ If $\mathbb{%
P}$ is the general Luce model induced by $(u,\Gamma ),$ then Con$(\mathbb{P}%
)=(0,1].$ It follows from Theorem 3.1 that $\mathbb{P}$ is minimally
rational. $\square $

\subsubsection{The 2-Stage Luce Model}

The famous Luce rule is maximally rational, but this is only the tip of the
iceberg. Let $u:X\rightarrow (0,\infty )$ be an arbitrary function and $%
\succcurlyeq $ a partial order on $X.$ The map $\mathbb{P}:X\times \mathfrak{%
X}\rightarrow \lbrack 0,1]$ defined by%
\begin{equation*}
\mathbb{P}(x,S):=\frac{u(x)}{\sum\limits_{\omega \in \text{MAX}%
(S,\succcurlyeq )}u(\omega )}\hspace{0.2in}\text{for any }x\in \text{%
{\footnotesize \textbf{MAX}}}(S,\succcurlyeq ),
\end{equation*}%
and $\mathbb{P}(x,S):=0$ for any $x\in X\backslash ${\footnotesize \textbf{%
MAX}}$(S,\succcurlyeq ),$ is a stochastic choice function on $\mathfrak{X}.$
This map is said to be the \textbf{2-stage} \textbf{Luce model} induced by $%
(u,\succcurlyeq )$. This model presumes that the agent has a certain type of
dominance relation $\succcurlyeq $ in her mind, and she does not pay any
attention to those alternatives in a menu that are dominated in terms of
this relation. In the context of any feasible menu, she first contracts the
menu by eliminating the dominated alternatives, and then chooses in
accordance with the Luce model from the resulting submenu. (If $\succcurlyeq 
$ is the equality relation, then the model reduces to the classical Luce
model.) Given this interpretation, it is only natural that $u$ and $%
\succcurlyeq $ are consistent. We thus say that this model is \textbf{proper}
if $u$ is \textit{increasing with respect to} $\succ $, that is, $u(x)>u(y)$
whenever $x\succ y$.\footnote{%
The class of all 2-stage Luce models is also characterized by Echenique and
Saito (2019). In addition, Horan (2021) has characterized those members of
this class whose dominance relations are (certain types of) semiorders on $X$%
. All members of the class characterized by Horan (2021) are proper.} The
upshot is that any such stochastic choice function is at least as rational
as \textit{any} stochastic choice function:

\bigskip

\noindent \textsc{Proposition 5.5.} \textit{Any proper 2-stage Luce model is
maximally rational.}

\smallskip 

\noindent \textsc{Proof. }See Appendix.

\bigskip 

Since some proper 2-stage Luce models are not RUM, Propositions 5.3 and 5.5
show that RUM is neither necessary nor sufficient for maximal rationality.

\subsection{Tremble Models of Choice}

In tremble models of choice, the random choice behavior of an individual is
viewed as arising from potential mistakes. In the simplest version of such
models, the individual is assumed to have a well-defined injective utility
function $u$ on $X$ that represents their deterministic preferences. With
probability $\alpha \in (\frac{1}{2},1),$ they choose the best alternative
in a menu $S$ relative to this order. And with probability $1-\alpha ,$ they
make a mistake -- their hand \textquotedblleft tremble\textquotedblright\ in
the moment of choice -- and as a result, they may choose any option in that
menu with equal probabilities. Formally, then, we say that a stochastic
choice function $\mathbb{P}\in $ \textsf{\textbf{scf}}$(X)$ is a \textbf{%
tremble model} induced by $(\alpha ,u)$ if%
\begin{equation*}
\mathbb{P}(x,S)=\alpha \mathbf{1}_{\arg \max u(S)}+(1-\alpha )\tfrac{1}{%
\left\vert S\right\vert }
\end{equation*}%
for every $S\in \mathfrak{X}$ and $x\in S.\footnote{%
To the best of our knowledge, these sorts of stochastic choice functions
were considered first by Harless and Camerer (1994).}$ We refer to $\alpha $
here as the \textbf{tremble probability} of the model.

Let $\mathbb{P}_{\alpha }$ be a tremble model induced by $(\alpha ,u).$
Obviously, $\mathbb{P}_{0}$ and $\mathbb{P}_{1}$ are maximally rational
(Examples 3.1 and 3.2), so we consider the case $0<\alpha <1.$ As $\mathbb{P}%
_{\alpha }$ is selective with respect to expansions, Ch$(\mathbb{P}_{\alpha
})=\varnothing $ (Lemma 3.6), and as it is also moderately s-transitive, ST$(%
\mathbb{P}_{\alpha })=\varnothing $ (Lemma 3.4). Next, take any $S\in 
\mathfrak{X}$ and $x\in S$, and denote the interval $(\mathbb{P}_{\alpha
}^{\ast }(x,S),\min_{y\in S}\mathbb{P}_{\alpha }^{\ast }(x,\{x,y\})]$ by $I$%
. If $u(x)>u(y)$ for every $y\in S,$ the left endpoint of $I$ is 1, so $%
I=\varnothing $. Assume, then, $u(y)>u(x)$ for some $y\in S.$ Then, a quick
calculation shows that $I$ equals $(\frac{1-\alpha }{1+(\left\vert
S\right\vert -1)\alpha },\frac{1-\alpha }{1+\alpha }].$ As the left endpoint
of this interval is decreasing in the size of $S,$ the smallest left
endpoint that can be obtained here is $\frac{1-\alpha }{1+(k-1)\alpha }$,
where $k:=\left\vert X\right\vert $. In view of Theorem 3.2, therefore, we
conclude that $\Lambda (\mathbb{P}_{\alpha })=(\frac{1-\alpha }{%
1+(k-1)\alpha },\frac{1-\alpha }{1+\alpha }].$

Now, $\alpha \mapsto \frac{1-\alpha }{1+\alpha }$ and $\alpha \mapsto \frac{%
1-\alpha }{1+(k-1)\alpha }$ are strictly decreasing maps on $(0,1).$ Thus, $%
\mathbb{P}_{\alpha }$ and $\mathbb{P}_{\beta }$ are not $\trianglerighteq _{%
\text{rat}}$-comparable for any distinct $\alpha ,\beta \in (0,1)$, in stark
contrast to uniform dRUM (Corollary 5.2.a). To examine the comparative
rationality of such cases, we thus use the rationality index $I_{\text{rat}}$%
. One can show that $\alpha \mapsto I_{\text{rat}}(\mathbb{P}_{\alpha })=1-%
\frac{1-\alpha }{1+\alpha }\frac{(k-2)\alpha }{(1+(k-1)\alpha )}$ is a
strictly convex map on $[0,1]$ whose global maxima are reached at $0$ and $1$
-- we have $I_{\text{rat}}(\mathbb{P}_{0})=1=I_{\text{rat}}(\mathbb{P}_{1})$
-- and whose global minimum is attained within $(0,1)$. Thus, our
rationality index $I_{\text{rat}}$ says: two tremble models with
sufficiently small (resp., large) tremble probabilities, the one with the
smaller (resp., larger) tremble probability is more rational.

In summary:

\bigskip

\noindent \textsc{Proposition 5.6.} \textit{a. No two tremble models with
distinct tremble probabilities are }$\trianglerighteq _{\text{rat}}$\textit{%
-comparable;}

\textit{b. There is a threshold }$0<\alpha _{0}<\frac{1}{2}$\textit{\ such
that any tremble model on} $\mathfrak{X}$\textit{\ whose tremble probability
is smaller (larger) than }$\alpha _{0}$\textit{\ becomes more rational
according to }$I_{\text{rat}}$\textit{\ as its tremble probability decreases
(increases).}

\bigskip

If we interpret trembles as \textquotedblleft mistakes\textquotedblright\
that are committed by a utility-maximizer at the time of choice, the choice
behavior should genuinely reflect its main source, namely, utility
maximization, and thus appear more rational as the likelihood of the
trembles gets smaller. In the limit where this likelihood is zero, the model
becomes the classical utility-maximization model, which is, of course,
maximally rational. Proposition 5.6 captures this idea, but also warns us
that this intuition is valid only when the tremble probabilities are small
enough, and overturns the conclusion when these probabilities are large
enough. On closer scrutiny, this too is intuitive. If $\alpha $ is small,
there is no reason to view trembles as \textquotedblleft
mistakes.\textquotedblright\ In that case, randomizing across all choice
options in a menu with equal probabilities is the main impetus behind the
agent's choices, the maximization of $u$ is an afterthought that happens
infrequently. Since choosing everything in every menu with equal
probabilities is a maximally rational behavior (Example 3.2) -- this, after
all, may well correspond to the behavior of a utility-maximizer who is
indifferent between all alternatives in $X$ -- in this case increasing the
tremble probabilities makes the model more rational, and in the limit (when $%
\alpha =0$), we arrive at the said maximally rational stochastic choice
model. It is worth noting, however, that these conclusions cannot be
sustained in terms of our rationality ordering $\trianglerighteq _{\text{rat}%
}$. The first part of the proposition above says that $\trianglerighteq _{%
\text{rat}}$ is unable to compare such stochastic choice functions; instead
the said intuition is delivered by the index $I_{\text{rat}}$.

\subsection{Empirical Applications}

The purpose of this section is to illustrate how one can use our measurement
methodology in the context of within-subject choice experiments.

\subsubsection{Experiments of Tversky (1969) and Regenwetter et. al. (2011)}

We first consider Experiment 1 of Tversky (1969) which was originally
designed for testing weak stochastic transitivity. The experiment recorded
the choices of each subject from all pairs of 5 gambles, where each trial
was repeated 20 times (with additional `decoy' choices in between), thereby
generating a sample stochastic choice data. Originally, this experiment had
only 8 subjects, but the same experiment was later replicated by Regenwetter
et al. (2011) with 18 additional subjects.\footnote{%
The only differences in the replication are the use of computers and updated
payoffs (as decades have passed), and the implementation of the experiment
in only one session (as opposed to Tversky's five sessions).} As in
Balakrishnan, Ok, and Ortoleva (2021), we pool these two data sets here to
obtain a choice data set of 26 individuals (with Subjects 19-26 coming from
Tversky's original experiment). As this is only an illustration, however, we
ignore sampling errors, and identify a subject's relative choice frequencies
at each pairwise menu with her choice probabilities at that menu.

\begin{figure}
\centering
\includegraphics[width=120mm]{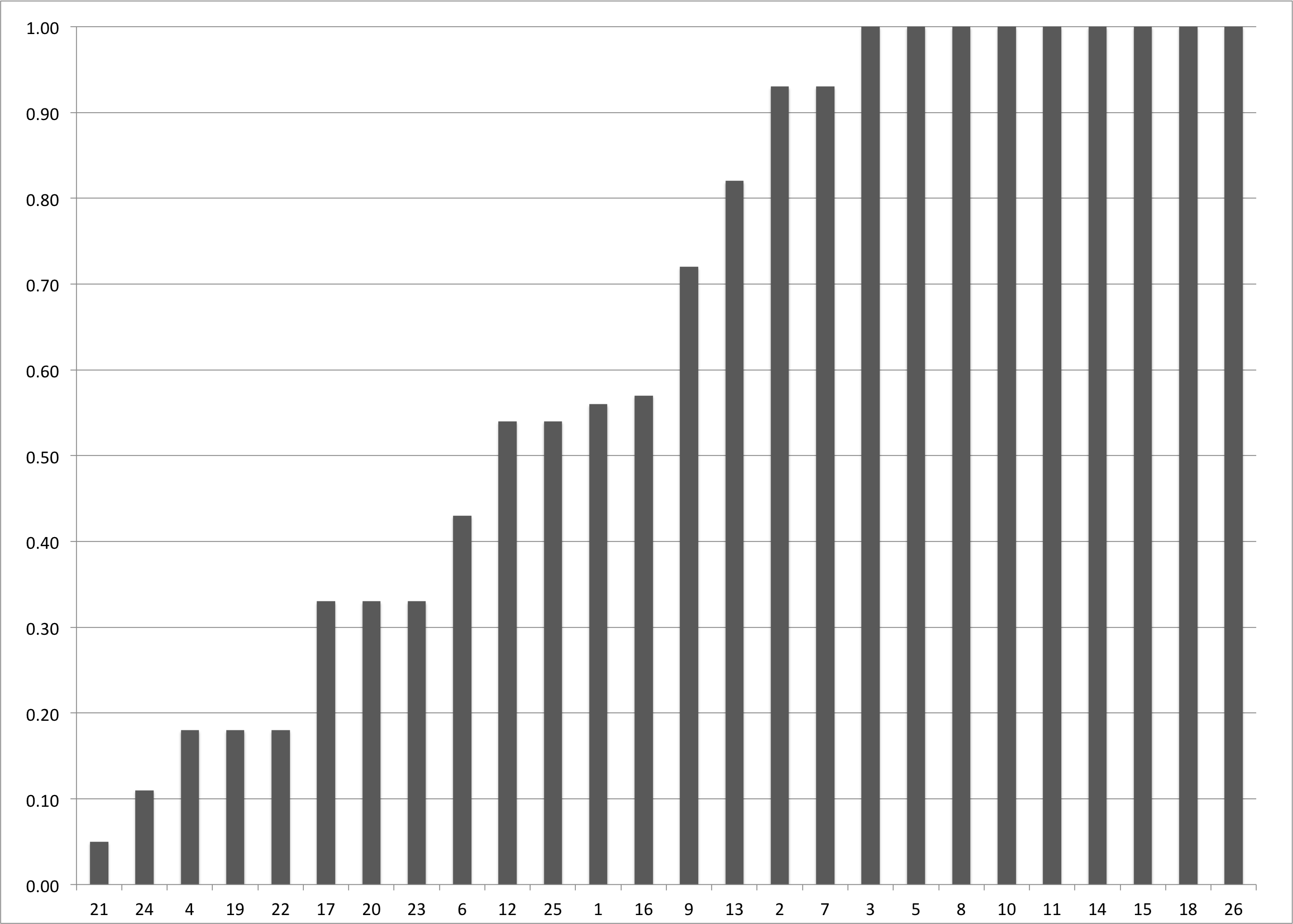}
\caption{Value Distribution of the Rationality Index}
\end{figure}

Figure 1 reports the values of our rationality index $I_{\text{rat}}$ for
all 26 subjects in the increasing order. (For instance, Subject 21 has the
lowest rationality, and Subjects 3, 5, $\ldots $, 26 have the highest,
according to $I_{\text{rat}}$.) We find that 9 out of 26 subjects are
maximally rational; i.e., their rationality index is equal to $1$.\footnote{%
By contrast, 12 subjects are weakly s-rational, and 8 of them are moderately
s-rational.} Moreover, $I_{\text{rat}}$ values of Subjects $2$ and $7$ (who
are also weakly stochastically transitive) are quite close to $1$ (both
equal to $0.93$), so if we were to allow for sampling errors (and put $I_{%
\text{rat}}(\mathbb{P})=1$ as a null hypothesis), we would have declared
these subjects too as maximally rational. In fact, an immediate perusal of
Figure 1 readily shows that there is a surprising amount of rationality in
the overall group, at least insofar as this is measured by $I_{\text{rat}}$.
Having said this, there are five subjects who fall way off the mark of
rationality. (For instance, $I_{\text{rat}}$ values of the Subjects 21, 24
and 4 are less than $0.2$.)

Let us now move to making rationality comparisons between subjects by using
the ordering $\trianglerighteq _{\text{rat}}$. Considering its
incompleteness, it was not clear \textit{a priori} if $\trianglerighteq _{%
\text{rat}}$ would yield any nontrivial comparisons. We were somewhat
surprised to find that it does. Given the pairwise nature of data (Section
3.4), $\trianglerighteq _{\text{rat}}$ here only depends on stochastic
transitivity sets. We have thus calculated these sets for each subject.
These are displayed in Figure 2 in the order of their rationality indices.
(Since the stochastic transitivity sets of the $9$ maximally rational
subjects are empty, we only display those of the remaining 17 subjects.)%

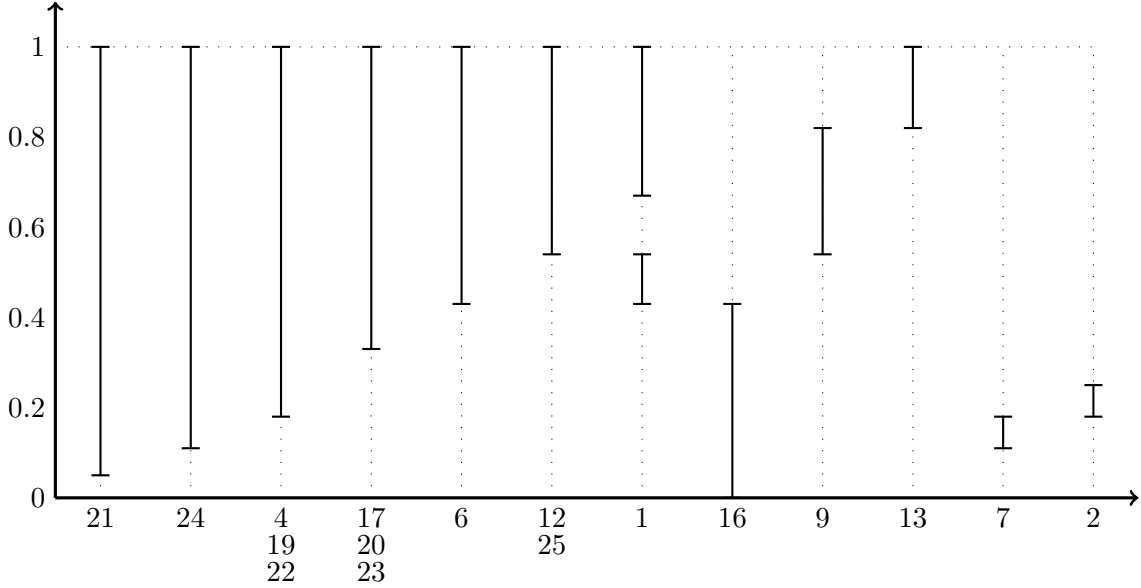
\begin{figure}
\begin{center}
\begin{tikzpicture}[x=0.60cm, y=0.60cm][domain=0:1, range=0:1, scale=3/4, thick]%paper
\usetikzlibrary{intersections}
\setlength{\unitlength}{0.5 cm}
%utility space
\draw[very thick][->] (-7,0)--(-7, 11) node[rotate=90, anchor=south east]{\text{}};
\draw[very thick][->] (-7,0)--(17,0) node[anchor=north east]{};
%\draw[->] (0,0)--(6,6);

\draw[loosely dotted] (-7,10)--(16,10);

\draw[loosely dotted] (-6,0)--(-6,10);

\draw[thick] (16,1.8)--(16,2.5);

\draw[thick] (15.8,2.5)--(16.2,2.5);
\draw[thick] (15.8,1.8)--(16.2,1.8);

\draw[loosely dotted] (-4,0)--(-4,10);

\draw[thick] (14,1.1)--(14,1.8);

\draw[thick] (13.8,1.8)--(14.2,1.8);
\draw[thick] (13.8,1.1)--(14.2,1.1);

\draw[loosely dotted] (-2,0)--(-2,10);

\draw[thick] (12,8.2)--(12,10);

\draw[thick] (11.8, 10)--(12.2,10);
\draw[thick] (11.8,8.2)--(12.2,8.2);

\draw[loosely dotted] (0,0)--(0,10);

\draw[thick] (10,5.4)--(10,8.2);

\draw[thick] (9.8,8.2)--(10.2,8.2);
\draw[thick] (9.8,5.4)--(10.2,5.4);

\draw[loosely dotted] (2,0)--(2,10);

\draw[thick] (8,0)--(8,4.3);

\draw[thick] (7.8,0)--(8.2,0);
\draw[thick] (7.8,4.3)--(8.2,4.3);

\draw[loosely dotted] (4,0)--(4,10);

\draw[thick] (6,4.3)--(6, 5.4);
\draw[thick] (6,6.7)--(6, 10);

\draw[thick] (5.8,10)--(6.2,10);
\draw[thick] (5.8,6.7)--(6.2,6.7);

\draw[thick] (5.8,5.4)--(6.2,5.4);
\draw[thick] (5.8,4.3)--(6.2,4.3);

\draw[loosely dotted] (6,0)--(6,10);

\draw[thick] (4,5.4)--(4,10);

\draw[thick] (3.8,10)--(4.2,10);
\draw[thick] (3.8,5.4)--(4.2,5.4);

\draw[loosely dotted] (8,0)--(8,10);

\draw[thick] (2,4.3)--(2,10);

\draw[thick] (1.8,10)--(2.2,10);
\draw[thick] (1.8,4.3)--(2.2,4.3);

\draw[loosely dotted] (10,0)--(10,10);

\draw[thick] (0,3.3)--(0,10);

\draw[thick] (-0.2,10)--(0.2,10);
\draw[thick] (-0.2,3.3)--(0.2,3.3);

\draw[loosely dotted] (12,0)--(12,10);

\draw[thick] (-2,1.8)--(-2,10);

\draw[thick] (-1.8,10)--(-2.2,10);
\draw[thick] (-1.8,1.8)--(-2.2,1.8);

\draw[loosely dotted] (14,0)--(14,10);

\draw[thick] (-4,1.1)--(-4,10);

\draw[thick] (-3.8,10)--(-4.2,10);
\draw[thick] (-3.8,1.1)--(-4.2,1.1);

\draw[loosely dotted] (16,0)--(16,10);

\draw[thick] (-6,0.5)--(-6,10);

\draw[thick] (-5.8,10)--(-6.2,10);
\draw[thick] (-5.8,0.5)--(-6.2,0.5);

%\filldraw[] (6,2) circle (3pt);
%\filldraw[] (4,4) circle (3pt);
%\filldraw[] (2,6) circle (3pt);

\coordinate[label=below:\text{21}] (wf) at (-6, 0);
\coordinate[label=below:\text{24}] (wf) at (-4, 0);
\coordinate[label=below:\text{4}] (wf) at (-2, 0);
\coordinate[label=below:\text{19}] (wf) at (-2, -0.6);
\coordinate[label=below:\text{22}] (wf) at (-2, -1.2);
\coordinate[label=below:\text{17}] (wf) at (0, 0);
\coordinate[label=below:\text{20}] (wf) at (0, -0.6);
\coordinate[label=below:\text{23}] (wf) at (0, -1.2);
\coordinate[label=below:\text{6}] (wf) at (2, 0);
\coordinate[label=below:\text{12}] (wf) at (4, 0);
\coordinate[label=below:\text{25}] (wf) at (4, -0.6);
\coordinate[label=below:\text{1}] (wf) at (6, 0);
\coordinate[label=below:\text{16}] (wf) at (8, 0);
\coordinate[label=below:\text{9}] (wf) at (10, 0);
\coordinate[label=below:\text{13}] (wf) at (12, 0);
\coordinate[label=below:\text{7}] (wf) at (14, 0);
\coordinate[label=below:\text{2}] (wf) at (16, 0);

\coordinate[label=left:$0$] (wf) at (-7, 0);
\coordinate[label=left:$0.2$] (wf) at (-7, 2);
\coordinate[label=left:$0.4$] (wf) at (-7, 4);
\coordinate[label=left:$0.6$] (wf) at (-7, 6);
%\coordinate[label=left:$0.8$] (wf) at (0, 8);
\coordinate[label=left:$0.8$] (wf) at (-7, 8);
\coordinate[label=left:$1$] (wf) at (-7, 10);

%\coordinate[label=above right: ${\bf{x}}$] (wf) at (6, 2);
%\coordinate[label=above right:${\bf{y}}$] (wf) at (4, 4);
%\coordinate[label=above right:${\bf{z}}$] (wf) at (2, 6);

\end{tikzpicture}
\end{center}
\caption{Distribution of $\Lambda$-sets for Tversky (1969)}
\end{figure}

Let $\mathbb{P}_{i}$ stand for the stochastic choice function for Subject $i$%
. One major highlight of Figure 2 is that the rationality of many subjects
can be compared by means of $\trianglerighteq _{\text{rat}}$. In fact, every
subject is either more or less rational than at least one other subject. For
example, $\Lambda (\mathbb{P}_{21})=(0.05,1]$ and Subject 21 is less
rational than all other subjects except Subject 16, that is, $\mathbb{P}%
_{i}\trianglerighteq _{\text{rat}}\mathbb{P}_{21}$ for every $i$ distinct
from 16, while Subject 16 is less rational than Subjects 2 and 7. Moreover, $%
\Lambda (\mathbb{P}_{1})=(0.43,0.54]\cup (0.67,1]$. So, Subject 1 is less
rational than Subject 13 but they are more rational than all the subjects
whose $I_{\text{rat}}$ values are strictly less than that of Subject 12.
(For instance, $\mathbb{P}_{1}\trianglerighteq _{\text{rat}}\mathbb{P}_{6}$%
.) Also evident from Figure 2 is that there are three groups of equally
rational subjects. For example, Subjects 12 and 15 have identical stochastic
transitivity sets. As a final observation of interest, we note that Subjects
2 and 7 have the same rationality index, and yet they are not comparable by $%
\trianglerighteq _{\text{rat}}$.

Standard tests of preference maximization, or stochastic transitivity, etc.,
are dichotomous: the choice data of a subject pass the adopted test or they
do not. However, one is often interested in the degree to which rationality is
violated. (To wit, Afriat's efficiency index and subsequent rationality
indices in the deterministic context of consumer theory were developed to
measure the severity of a violation of the generalized axiom of revealed
preference.) A major advantage of our rationality index $I_{\text{rat}}$ and
rationality order $\trianglerighteq _{\text{rat}}$ stems from the fact that
they allow the analyst to measure the extent of rationality violations in
any stochastic choice environment. For example, Tversky concluded in his
original study that Subjects 21 and 25 both violate weak stochastic
transitivity (even accounting for sampling errors). $I_{\text{rat}}$ and $%
\trianglerighteq _{\text{rat}}$ provide more information about the situation
by showing that Subject 21 is, in fact, far less rational than Subject 25.

In passing, it may be useful to contrast our approach to that of evaluating
this experimental choice data by using the RUM. In general, testing the RUM
on binary choices is notoriously difficult. (See Sprumont (2022) for a
recent discussion of this issue.) Fortunately, when the total number of
alternatives is at most 5, that is, $\left\vert X\right\vert \leq 5,$
Theorem 1 of Fishburn (1992) establishes that we only need to check the
following \textit{triangular condition} to this effect:%
\begin{equation*}
\mathbb{P}(x,\{x,y\})+\mathbb{P}(y,\{y,z\})+\mathbb{P}(z,\{x,z\})\leq 2
\end{equation*}%
for every $x,y,z\in X.$ It turns out that maximal rationality implies this
condition. Consequently, we can readily conclude that the 9 maximally
rational subjects of the experiment are sure to be consistent with the RUM.
In addition to these, using Fishburn's condition directly, we find that the
subjects 2, 7, 13, 9, 1, 12, and 25 are also consistent with the RUM. It is
worth noting that four of these subjects, namely, S13, S1, S12 and S25,
violate weak s-transitivity. This is but an empirical testament to the fact
that a RUM need not be weakly s-transitive. Moreover, it shows that the RUM
cannot distinguish the choice behavior of S1, S12, and S25 from those of S2,
S7, and S9 who satisfy weak s-transitivity. In contrast, as we have seen in
Figure 2, the present method of measuring stochastic rationality renders
sharper comparisons. For instance, it follows from Figure 2 that any one of
S2, S7, and S9 is more rational than S1, S12 and S25 with respect to $I_{%
\text{rat}}$. In fact, S9 is more rational than S12 and S25 even relative to 
$\trianglerighteq _{\text{rat}}$.

\subsubsection{Experiments of Feldman and Rehbeck (2022)}

To demonstrate the computational feasibility of our algorithm for
calculating $\Lambda (\mathbb{P})$, we next consider the experiments of
Feldman and Rehbeck (2022). In these experiments, the subjects interacted
with a linear convex budget interface in the context of risky choice. The
authors found that out of 144 subjects, 56 subjects made non-degenerate
random choices in more than half of the 79 menus they faced, while 36 of
them did so for more than two-thirds of the menus.

Using the algorithm we introduced in Section 4.2, it took us less than ten
minutes on a personal computer to calculate all the acyclicity sets for the
entirety of the subjects. We found that only 1 subject was maximally
rational, while 122 subjects were minimally rational. In most of the
existing experimental studies, the subjects face at most 25 menus. For
example, in Tversky (1969), subjects were confronted with only 10 menus, and
in the convex budget experiments of Choi et al. (2007) and further studies
inspired by it, the subjects faced between 20 to 25 menus. As the values of
rationality indices are naturally decreasing in the number of menus, it is
not surprising that the overall rationality of the subjects in Feldman and
Rehbeck (2022) is lower than that in other experimental studies. However,
there are other potential reasons why we observe such a low level of
rationality. If a subject's choices are deterministic, then our rationality
order would conclude that they are either maximally or minimally rational.
In fact, most of the 122 minimally rational subjects tended to choose
degenerate randomizations in the experiment, which resulted in drastic
departures from rationality. To wit, Subject 1's choice exhibits the
following deterministic cycle: $\mathbb{P}(a_{2},\{a_{2},a_{11}\})=\mathbb{P}%
(a_{11},\{a_{1},a_{11}\})=\mathbb{P}(a_{1},\{a_{1},a_{12}\})=1$ and $\mathbb{%
P}(a_{2},\{a_{2},a_{12}\})=0$ where alternatives are denoted as $%
a_{1},\ldots ,a_{27}$. Subjects who have intermediate levels of rationality
tend to randomize more. For 19 out of 21 subjects who were neither minimally
nor maximally rational, more than two-thirds of their choices were random.
(This is the property of the data, not of our measure.) 

Still, a good fraction of subjects who exhibited random choice were minimally
rational. This seems consistent with the experimental findings of Nielsen
and Rigotti (2023) who propose a novel method to detect incompleteness of
preferences and find that many subjects in
their experiment exhibit incompleteness.\footnote{Note that many economists argue that there is a deeper connection between incompleteness of preferences and random choice, e.g., Ok and Tserenjigmid (2022) and Agranov and Ortoleva (2023).} It is telling that Nielsen and
Rigotti observed that when their subjects are forced to make choices (as in
most existing experimental studies, and certainly in that of Feldman and
Rehbeck (2022)), the observed choices exhibit more inconsistencies compared
to non-forced situations.%

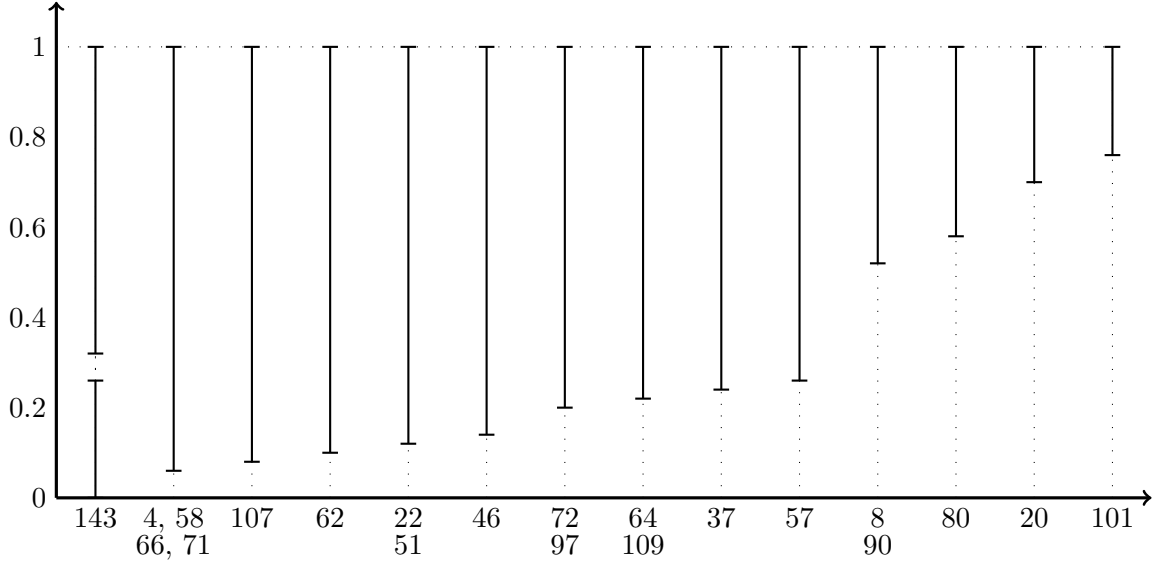
\begin{figure}
\begin{center}
\begin{tikzpicture}[x=0.52cm, y=0.60cm][domain=0:1, range=0:1, scale=3/4, thick]%paper
\usetikzlibrary{intersections}
\setlength{\unitlength}{0.5 cm}
%utility space
\draw[very thick][->] (-7,0)--(-7, 11) node[rotate=90, anchor=south east]{\text{}};
\draw[very thick][->] (-7,0)--(21,0) node[anchor=north east]{};
%\draw[->] (0,0)--(6,6);

\draw[loosely dotted] (-7,10)--(20,10);

\draw[loosely dotted] (20,0)--(20,10);

\draw[loosely dotted] (-6,0)--(-6,10);

\draw[thick] (20,10)--(20,7.6);

\draw[thick] (19.8,10)--(20.2,10);

\draw[thick] (19.8,7.6)--(20.2,7.6);

\draw[thick] (18,10)--(18,7);

\draw[thick] (17.8,10)--(18.2,10);

\draw[thick] (17.8,7)--(18.2,7);

\draw[loosely dotted] (18,0)--(18,7);

\draw[thick] (16,10)--(16,5.8);

\draw[thick] (15.8,10)--(16.2,10);

\draw[thick] (15.8,5.8)--(16.2,5.8);

\draw[loosely dotted] (16,0)--(16,5.8);

\draw[thick] (14,10)--(14,5.2);

\draw[thick] (13.8,10)--(14.2,10);

\draw[thick] (13.8,5.2)--(14.2,5.2);

\draw[loosely dotted] (14,0)--(14,5.2);

\draw[thick] (12,10)--(12,2.6);

\draw[thick] (11.8,10)--(12.2,10);

\draw[thick] (11.8,2.6)--(12.2,2.6);

\draw[loosely dotted] (12, 0)--(12,2.6);

\draw[thick] (10,10)--(10,2.4);

\draw[thick] (9.8,10)--(10.2,10);

\draw[thick] (9.8,2.4)--(10.2,2.4);

\draw[loosely dotted] (10,0)--(10,2.4);

\draw[thick] (8,10)--(8,2.2);

\draw[thick] (7.8,10)--(8.2,10);

\draw[thick] (7.8,2.2)--(8.2,2.2);

\draw[loosely dotted] (8,0)--(8,2.2);

\draw[thick] (6,10)--(6,2);

\draw[thick] (5.8,10)--(6.2,10);

\draw[thick] (5.8,2)--(6.2,2);

\draw[loosely dotted] (6,0)--(6,2);

\draw[thick] (4,10)--(4,1.4);

\draw[thick] (3.8,10)--(4.2,10);

\draw[thick] (3.8,1.4)--(4.2,1.4);

\draw[loosely dotted] (4,0)--(4,1.4);

\draw[thick] (2,10)--(2,1.2);

\draw[thick] (1.8,10)--(2.2,10);

\draw[thick] (1.8,1.2)--(2.2,1.2);

\draw[loosely dotted] (2,0)--(2,1.2);

\draw[thick] (0,10)--(0,1);

\draw[thick] (-0.2,10)--(0.2,10);

\draw[thick] (-0.2,1)--(0.2,1);

\draw[loosely dotted] (0,0)--(0,1);

\draw[thick] (-2,10)--(-2,0.8);

\draw[thick] (-2.2,10)--(-1.8,10);

\draw[thick] (-2.2,0.8)--(-1.8,0.8);

\draw[loosely dotted] (-2,0)--(-2,0.8);

\draw[thick] (-4,10)--(-4,0.6);

\draw[thick] (-4.2,10)--(-3.8,10);

\draw[thick] (-4.2,0.6)--(-3.8,0.6);

\draw[loosely dotted] (-4,0)--(-4,0.6);

\draw[thick] (-6,10)--(-6,3.2);

\draw[thick] (-6.2,10)--(-5.8,10);

\draw[thick] (-6.2,3.2)--(-5.8,3.2);

\draw[thick] (-6,0)--(-6,2.6);

\draw[thick] (-6.2,2.6)--(-5.8,2.6);

\draw[thick] (-6.2,0)--(-5.8,0);

\draw[loosely dotted] (-6,2.6)--(-6,3.2);

\coordinate[label=below:\text{143}] (wf) at (-6, 0); %%49
\coordinate[label=below:\text{4, 58}] (wf) at (-4, 0); %%47
%\coordinate[label=below:\text{58}] (wf) at (-4, -0.6); %%47
\coordinate[label=below:\text{66, 71}] (wf) at (-4, -0.6); %%47
%\coordinate[label=below:\text{71}] (wf) at (-4, -1.8); %%47
\coordinate[label=below:\text{107}] (wf) at (-2, 0); %%46
%\coordinate[label=below:\text{}] (wf) at (-2, -0.6);
%\coordinate[label=below:\text{22}] (wf) at (-2, -1.2);
\coordinate[label=below:\text{62}] (wf) at (0, 0); %%45
%\coordinate[label=below:\text{20}] (wf) at (0, -0.6);
%\coordinate[label=below:\text{23}] (wf) at (0, -1.2);
\coordinate[label=below:\text{22}] (wf) at (2, 0); %%44
\coordinate[label=below:\text{51}] (wf) at (2, -0.6); %%44
%\coordinate[label=below:\text{62}] (wf) at (2, -1.2); 
\coordinate[label=below:\text{46}] (wf) at (4, 0); %%43
%\coordinate[label=below:\text{25}] (wf) at (4, -0.6);
\coordinate[label=below:\text{72}] (wf) at (6, 0); %%40
\coordinate[label=below:\text{97}] (wf) at (6, -0.6); %%40
\coordinate[label=below:\text{64}] (wf) at (8, 0); %%39
\coordinate[label=below:\text{109}] (wf) at (8, -0.6); %%39
\coordinate[label=below:\text{37}] (wf) at (10, 0); %%38
\coordinate[label=below:\text{57}] (wf) at (12, 0); %%37
\coordinate[label=below:\text{8}] (wf) at (14, 0); %%24
\coordinate[label=below:\text{90}] (wf) at (14, -0.6); %%24
\coordinate[label=below:\text{80}] (wf) at (16, 0);%%21
\coordinate[label=below:\text{20}] (wf) at (18, 0);%%15
\coordinate[label=below:\text{101}] (wf) at (20, 0);%%12

\coordinate[label=left:$0$] (wf) at (-7, 0);
\coordinate[label=left:$0.2$] (wf) at (-7, 2);
\coordinate[label=left:$0.4$] (wf) at (-7, 4);
\coordinate[label=left:$0.6$] (wf) at (-7, 6);
%\coordinate[label=left:$0.8$] (wf) at (0, 8);
\coordinate[label=left:$0.8$] (wf) at (-7, 8);
\coordinate[label=left:$1$] (wf) at (-7, 10);

%\coordinate[label=above right: ${\bf{x}}$] (wf) at (6, 2);
%\coordinate[label=above right:${\bf{y}}$] (wf) at (4, 4);
%\coordinate[label=above right:${\bf{z}}$] (wf) at (2, 6);

\end{tikzpicture}
\end{center}
\caption{Distribution of $\Lambda$-sets for Feldman and Rehbeck (2022)}
\end{figure}

Figure 3 depicts the acyclicity sets of the 21 subjects whose levels of
rationality are not extreme; we ordered these subjects from lowest to
highest according to their $I_{\text{rat}}$ values. (For instance, Subject
143 has the lowest rationality, and Subject 101 has the highest, according
to $I_{\text{rat}}$.) The acyclicity set of Subject 143 is $(0,0.3]\cup
(0.32,1]$, and for any other subjects, their acyclicity sets are of the form 
$(\underline{\lambda },1]$ of a single interval containing $1$. Hence,
except Subject 143, any two subjects are comparable by $\trianglerighteq _{%
\text{rat}}$; in the case of these participants the calls of $%
\trianglerighteq _{\text{rat}}$ and $I_{\text{rat}}$ are almost identical.
We thus observe that $\trianglerighteq _{\text{rat}}$ is able to make many
comparisons in the present data set as well. 

%\begin{equation*}
%\text{FIGURE: Value Distribution}
%\end{equation*}

\section{Conclusion}

The great majority of economic models presume that individuals make their
choices in a rational manner. Depending on the context, the term
\textquotedblleft rational\textquotedblright\ means different things, but
being a \textquotedblleft preference maximizer\textquotedblright\ with
respect to some transitive (and often complete) preference relation is
almost always part of the definition. Besides, choice theory provides
various ways of capturing this property from a behavioral viewpoint by means
of (testable) revealed preference axioms, and the recent literature has
discovered methods of making rationality comparisons between different
(deterministic) choice behaviors. However, when the individual choice
behavior is modeled as random, determining which models are rational becomes
a nontrivial matter. The literature seems to consider the random utility
model as the analogue of the standard utility maximization model, thereby
viewing that model as the canonical \textquotedblleft
rational\textquotedblright\ stochastic choice function. But as we have
discussed in Sections 1 and 5.1, this is not an unexceptionable viewpoint,
absent suitable consistency conditions among the utilities of the model.%
\footnote{\linespread{1.1} \selectfont One might think that deliberately
stochastic choice models (e.g., studied in Machina (1985), Fudenberg,
Iijima, and Strzalecki (2015), and Cerreia-Vioglio, Dillenberger, Ortoleva,
and Riella (2019)) are trivially maximally rational as in these models a
decision maker maximizes a preference over probability distributions over
alternatives. Deliberately stochastic choice models may or may not be
maximally rational by $\trianglerighteq _{\text{rat}}$ since, in general,
the preference over probability distributions and the preference over
alternatives are not required to be consistent with each other. Moreover, as
shown in Cerreia-Vioglio, Dillenberger, Ortoleva, and Riella (2019), some
deliberately stochastic choice models may violate weak stochastic
transitivity.}

In this paper, we have thus developed a rigorous method for comparing the
rationality of any two random choice models, may these be empirical or
theoretical, thereby producing criteria for maximal, as well as minimal,
rationality for such models. The basic idea is to \textquotedblleft
approximate\textquotedblright\ a given stochastic choice function with a
deterministic choice correspondence, and check whether or not that
correspondence is rational. This allows us to bring well-known (deterministic)
revealed preference methods to assess the rationality of the approximating
choice correspondence. But there is an immediate difficulty with this idea.
It seems impossible to agree on one particular approximation, as this may
simply cause too much loss of information. To alleviate this issue, we
considered here a one-parameter class of potential, and intuitively
appealing, approximations (the Fishburn family), and compared the
rationality of any two random choice models by using \textit{all} members of
this class. This has furnished us with a dominance ranking $\trianglerighteq
_{\text{rat}}$. We have then provided a parameter-free characterization of
this ranking, which led to a natural rationality index $I_{\text{rat}}$ that
is duly consistent with $\trianglerighteq _{\text{rat}}$.

Endowed with this methodology, we have revisited the random utility model,
and found (this time formally) that such a model may or may not be deemed
maximally rational by $\trianglerighteq _{\text{rat}}$, depending on the
structure of the utilities of the model. In the case where there are two
utility functions in play, we have found that $\trianglerighteq _{\text{rat}}
$ is able to rank all such models, and characterized the maximally rational
ones by means of a simple consistency condition. We have also applied our
measurement methodology to some other models such as the generalized Luce
model and the tremble model, and demonstrated how it can also be utilized
empirically by using the choice data generated by repeated, as well as
random, choice experiments.

There remain several avenues of research to explore. First and foremost,
this paper motivates studying the problem of approximating a given
stochastic choice function by a deterministic choice correspondence
seriously. The Fishburn family is an intuitive and convenient class, but it
is by no means the only option in this regard. One can/should approach the
general approximation problem either axiomatically or by means of metric
projection methods (e.g., along the lines of Apesteguia and Ballester
(2021)), and obtain other viable alternatives. Second, even adopting the
Fishburn class as is, one may study other methods of ranking choice
correspondences in order to obtain more refined rationality orderings. Along
with a host of others, these problems are left for future research.

\bigskip

\bigskip

%TCIMACRO{\TeXButton{TeX field}{\begin{small}}}%
%BeginExpansion
\begin{small}%
%EndExpansion

{\small \noindent }{\Large \textbf{Apppendix: Proofs}}

\smallskip

\smallskip

{\small \noindent \textsc{Proof of Proposition 3.3.} Take any} $x,y,z\in X$
such that $a:=\mathbb{P}(x,\{x,y\})>\frac{1}{2}$ and $b:=\mathbb{P}%
(y,\{y,z\})>\frac{1}{2}$. Put $c:=\mathbb{P}(x,\{x,z\});$ we need to show
that $c\geq \min \{a,b\}.$ Since $a>\frac{1}{2},$ we have $\frac{1-a}{a}<1,$
and similarly, $\frac{1-b}{b}<1.$ Pick any real number $\lambda $ with 
\begin{equation*}
\max \left\{ \frac{1-a}{a},\frac{1-b}{b}\right\} <\lambda \leq 1.
\end{equation*}%
Then, $1-a<\lambda a,$ that is, $\mathbb{P}(y,\{x,y\})<\lambda \mathbb{P}%
(x,\{x,y\}),$ and similarly, $\mathbb{P}(z,\{y,z\})<\lambda \mathbb{P}%
(y,\{y,z\})$. Since $\mathbb{P}$ is maximally rational, it is $\lambda $%
-stochastic transitive (Proposition 3.2), so it follows that $\mathbb{P}%
(z,\{x,z\})<\lambda \mathbb{P}(x,\{x,z\}),$ that is, $\frac{1-c}{c}<\lambda
. $ Letting $\lambda \downarrow \max \left\{ \frac{1-a}{a},\frac{1-b}{b}%
\right\} $ then yields%
\begin{equation*}
\frac{1-c}{c}\leq \max \left\{ \frac{1-a}{a},\frac{1-b}{b}\right\} \text{,}
\end{equation*}%
that is, either $\frac{1-c}{c}\leq \frac{1-a}{a}$ or $\frac{1-c}{c}\leq 
\frac{1-b}{b}$. Since $t\mapsto \frac{1-t}{t}$ is a strictly decreasing
function on $(0,1],$ we thus find $c\geq a$ or $c\geq b,$ as we sought.%
{\small \ \ $\blacksquare $}

\bigskip

{\small \noindent \textsc{Proof of Lemma 3.4.} Let $\mathbb{P}$ be a
moderately stochastically transitive stochastic choice function on $%
\mathfrak{X}$\ such that ST$(\mathbb{P})\neq \varnothing $. Then, there must
exist $x,y,z\in X$ such that 
\begin{equation}
\mathbb{P}^{\ast }(z,\{x,z\})>\mathbb{P}^{\ast }(y,\{x,y\})\hspace{0.2in}%
\text{and\hspace{0.2in}}\mathbb{P}^{\ast }(z,\{x,z\})>\mathbb{P}^{\ast
}(z,\{y,z\}).  \label{mst}
\end{equation}%
By the first of these inequalities, $\mathbb{P}^{\ast }(y,\{x,y\})<1,$ which
is possible iff $\mathbb{P}(x,\{x,y\})>\frac{1}{2}$. Similarly, the second
inequality of (\ref{mst}) implies $\mathbb{P}(y,\{y,z\})>\frac{1}{2}$. It
then follows from moderate stochastic transitivity that $\mathbb{P}%
(x,\{x,z\})>\frac{1}{2}$. Consequently, (\ref{mst}) becomes%
\begin{equation*}
\frac{1-\mathbb{P}(x,\{x,z\})}{\mathbb{P}(x,\{x,z\})}>\frac{1-\mathbb{P}%
(x,\{x,y\})}{\mathbb{P}(x,\{x,y\})}\hspace{0.2in}\text{and\hspace{0.2in}}%
\frac{1-\mathbb{P}(x,\{x,z\})}{\mathbb{P}(x,\{x,z\})}>\frac{1-\mathbb{P}%
(y,\{y,z\})}{\mathbb{P}(y,\{y,z\})},
\end{equation*}%
that is, $\mathbb{P}(x,\{x,y\})>\mathbb{P}(x,\{x,z\})$ and $\mathbb{P}%
(y,\{y,z\})>\mathbb{P}(x,\{x,z\})$. Since both $\mathbb{P}(x,\{x,y\})$ and $%
\mathbb{P}(y,\{y,z\})$ exceed $\frac{1}{2},$ this contradicts $\mathbb{P}$
being moderately stochastically transitive.\ $\blacksquare $ }

\bigskip

{\small \noindent \textsc{Proof of Proposition 5.1.} Let us write $I_{S,x}$
for the interval $(\mathbb{P}^{\ast }(x,S),\min_{y\in S}\mathbb{P}^{\ast
}(x,\{x,y\})]$ for any $S\in \mathfrak{X}$ and $x\in S$. We note that if $%
\theta >\frac{1}{2}$ and the right endpoint of such an interval $I_{S,x}$ is 
$1,$ then $I_{S,x}=\varnothing $. (Indeed, when $\theta >\frac{1}{2},$ the
right endpoint of $I_{S,x}$ is 1 only if $u(x)>u(y)$ for all $y\in S,$ which
means $\mathbb{P}^{\ast }(x,S)=1$, that is, $I_{S,x}=\varnothing $.) }

{\small Suppose that $u$ and $v$ are not consistent over triplets, so (\ref%
{con}) holds for some $x,y,z\in X.$ Then, for $T:=\{x,y,z\},$ we have $%
\mathbb{P}^{\ast }(y,T)=0$ while $\mathbb{P}^{\ast }(y,\{x,y\})=\frac{%
1-\theta }{\theta }$ and $\mathbb{P}^{\ast }(y,\{z,y\})=1,$ and it follows
that $I_{T,y}=(0,\frac{1-\theta }{\theta }]\subseteq $ Con$(\mathbb{P})$. If 
$\theta =\frac{1}{2},$ the converse containment holds obviously. If $\theta >%
\frac{1}{2},$ then as we have noted in the previous paragraph, for any $S\in 
\mathfrak{X}$ and $w\in S$, the right endpoint of $I_{S,w}$ cannot be 1,
unless that interval is empty. Since $\mathbb{P}^{\ast }$ is $\{0,\frac{%
1-\theta }{\theta },1\}$-valued, we may thus conclude: If $u$ and $v$ are
not consistent over triplets, then $\Lambda (\mathbb{P})=(0,\frac{1-\theta }{%
\theta }].$ }

{\small Assume next that $u$ and $v$ are consistent over triplets. Take any $%
S\in \mathfrak{X}$ and $x\in S$. To derive a contradiction, suppose $%
I:=I_{S,x}\neq \varnothing $ (which implies trivially that the right
endpoint of this interval is nonzero). As noted above, $\theta >\frac{1}{2}$
implies that the right endpoint of $I$ cannot be 1. It follows that $%
\min_{y\in S}\mathbb{P}^{\ast }(x,\{x,y\})=\frac{1-\theta }{\theta }$ when $%
\theta >\frac{1}{2}$. This conclusion is also true for $\theta =\frac{1}{2},$
simply because $\mathbb{P}^{\ast }$ is then $\{0,1\}$-valued. We may thus
conclude that $x$ does not maximize $u$ on $S$, so there exists a $y\in S$
with $u(y)>u(x).$ We must then have $v(x)>v(y),$ for otherwise the right
endpoint of $I$ is zero. Moreover, if $v(x)=\max v(S),$ then $\mathbb{P}%
^{\ast }(x,S)=\frac{1-\theta }{\theta },$ so $I=\varnothing $, a
contradiction. Otherwise, there is a $z\in S$ with $v(z)>v(x).$ If $%
u(z)>u(x),$ then $\mathbb{P}(x,\{x,z\})=0,$ so we contradict $\min_{y\in S}%
\mathbb{P}^{\ast }(x,\{x,y\})=\frac{1-\theta }{\theta }$. On the other hand,
if $u(x)>u(z),$ then $u$ and $v$ are not consistent over triplets, a
contradiction. Conclusion: If $u$ and $v$ are consistent over triplets, then 
$\Lambda (\mathbb{P})=\varnothing $, that is, $\mathbb{P}$ is maximally
rational.$\blacksquare $}

\bigskip

{\small \noindent \textsc{Proof of Proposition 5.3.} (a) Let $\frac{1}{2}%
\geq \theta _{1}$ and take any $x,x_{1},...,x_{n}\in X$ such that (\ref{conn}%
) holds. Then, $\mathbb{P}(x,\{x,x_{j}\})=1-\theta _{j}\geq \frac{1}{2}$,
whence $\mathbb{P}^{\ast }(x,\{x,x_{j}\})=1,$ for all $j\in N,$ while $%
\mathbb{P}^{\ast }(x,\{x,x_{1},...,x_{n}\})=0.$ It then follows from the
definition of the Condorcet set of $\mathbb{P}$ that Con$(\mathbb{P})=(0,1].$
Thus: $\Lambda (\mathbb{P})=(0,1].$ }

{\small (b) Let $\theta _{1}>\frac{1}{2}$ and take any $x,y,z\in X$ such
that (\ref{connn}) holds. Then, $\mathbb{P}^{\ast }(y,\{x,y\})=\frac{%
1-\theta _{1}}{\theta _{1}}$ and $\mathbb{P}^{\ast }(y,\{y,z\})=1$ while $%
\mathbb{P}^{\ast }(y,\{x,y,z\})=0.$ It then follows from the definition of
the Condorcet set of $\mathbb{P}$ that $(0,\frac{1-\theta _{1}}{\theta _{1}}%
]\subseteq $ Con$(\mathbb{P})\subseteq \Lambda (\mathbb{P}).$ }

{\small To streamline the rest of the argument, we define the measure $\mu $
on $2^{N}$ by $\mu (\mathfrak{\varnothing })=0$ and $\mu (A):=\sum_{i\in
A}\theta _{i}$. }

{\small \textit{Claim 1. }Ch$(\mathbb{P})\subseteq (0,\frac{1-\theta _{1}}{%
\theta _{1}}]$. To see this, take any $S,T\in \mathfrak{X}$ and $x\in S$
with $S\subseteq T$ and $\mathbb{P}^{\ast }(x,T)>\mathbb{P}^{\ast }(x,S).$
Put $A:=\{i\in N:u_{i}(x)=\max u_{i}(S)\}$ and $B:=\{i\in N:u_{i}(x)=\max
u_{i}(T)\}.$ Obviously, $B\subseteq A.$ Moreover, as $1>\mathbb{P}^{\ast
}(x,S),$ $x$ does not maximize $u_{1}$ on $S,$ so $\mu (A)\leq 1-\theta _{1}$%
. As $x$ does not maximize $u_{1}$ on $T$ (because $S\subseteq T$) and $%
\theta _{1}>\frac{1}{2}$, we also have $\max_{w\in S}\mathbb{P}(w,T)\geq
\theta _{1}$. But then 
\begin{equation*}
\mathbb{P}^{\ast }(x,T)\leq \tfrac{\mu (B)}{\theta _{1}}\leq \tfrac{\mu (A)}{%
\theta _{1}}\leq \tfrac{1-\theta _{1}}{\theta _{1}},
\end{equation*}%
and our claim follows from the definition of Ch$(\mathbb{P})$. }

{\small \textit{Claim 2.} Con$(\mathbb{P})\subseteq (0,\frac{1-\theta _{1}}{%
\theta _{1}}]$. To see this, take any $S\in \mathfrak{X}$ and $x\in S.$
Define $A$ as in the previous paragraph, and let $z$ be an alternative in $S$
with $\mathbb{P}(z,S)=\max_{w\in S}\mathbb{P}(w,S).$ We put $C:=\{i\in
N:u_{i}(z)=\max u_{i}(S)\}$ so that $\mathbb{P}(z,S)=\mu (C).$ Since $\theta
_{1}>\frac{1}{2},$ we have $1\in C.$ But then $\mathbb{P}^{\ast }(x,S)=\frac{%
\mu (A)}{\mu (C)}$ and 
\begin{equation*}
\min_{y\in S}\mathbb{P}^{\ast }(x,\{x,y\})\leq \mathbb{P}^{\ast
}(x,\{x,z\})\leq \tfrac{1-\mu (C)}{\mu (C)}\leq \tfrac{1-\theta _{1}}{\theta
_{1}},
\end{equation*}%
and it follows from the definition of the Condorcet set of $\mathbb{P}$ that
Con$(\mathbb{P})\subseteq (0,\frac{1-\theta _{1}}{\theta _{1}}].$ }

{\small \textit{Claim 3.} ST$(\mathbb{P})\subseteq (0,\frac{1-\theta _{1}}{%
\theta _{1}}]$. To see this, take any $x,y,z\in X$ such that $\mathbb{P}%
^{\ast }(y,\{x,y\})<\mathbb{P}^{\ast }(z,\{x,z\})$ and $\mathbb{P}^{\ast
}(z,\{y,z\})<\mathbb{P}^{\ast }(z,\{x,z\}).$ Then $u_{1}(y)>u_{1}(x)$ cannot
be the case, for otherwise $\mathbb{P}^{\ast }(y,\{x,y\})=1$. Similarly, $%
u_{1}(z)>u_{1}(y),$ for otherwise $\mathbb{P}^{\ast }(z,\{y,z\})=1$. Thus: $%
u_{1}(x)>u_{1}(y)>u_{1}(z).$ It follows that $\mathbb{P}(z,\{x,z\})\leq
1-\theta _{1},$ so $\mathbb{P}^{\ast }(z,\{x,z\})\leq \tfrac{1-\theta _{1}}{%
\theta _{1}}$, and our claim follows from the definition of ST$(\mathbb{P})$%
. }

{\small Claims 1-3 jointly imply $\Lambda (\mathbb{P})\subseteq (0,\frac{%
1-\theta _{1}}{\theta _{1}}],$ completing our proof. $\blacksquare $}

\bigskip

{\small \noindent \textsc{Proof of Proposition 5.4.} \textit{Claim 1. }Ch$(%
\mathbb{P})=\varnothing $. To see this, take any $S,T\in \mathfrak{X}$ and $%
x\in S$ with $S\subseteq T$. Suppose $I:=($$\mathbb{P}^{\ast }(x,S),\mathbb{P%
}^{\ast }(x,T)]$ is nonempty. Then, $u_{1}(x)<\max u_{1}(S),$ because
otherwise, $\mathbb{P}^{\ast }(x,S)=1$ (given that $\theta _{1}>\frac{1}{2}$%
). If there is a $z\in T$ with $u_{1}(y)=\max u_{1}(T)>u_{1}(z)>u_{1}(x)$,
then by hypothesis, $u_{i}(z)>u_{i}(x)$ for all $i\in N,$ whence $\mathbb{P}%
^{\ast }(x,T)=0,$ contradicting $I$ being nonempty. Assume, then, $%
u_{1}(y)>u_{1}(x)>u_{1}(z)$ for each $z\in T\setminus \{x,y\}$. Then, it
must be that $y\in S$. Moreover, by hypothesis, $u_{i}(y)>u_{i}(z)$ for all $%
i\in N$ and $z\in T\backslash \{y\}.$ Finally, it is easy to check that $%
\mathbb{P}(y,\{x,y\})=\mathbb{P}(y,S)=\mathbb{P}(y,T)$ and $\mathbb{P}%
(x,\{x,y\})=\mathbb{P}(x,S)=\mathbb{P}(x,T)$, which means $I$ is empty, a
contradiction. Our claim thus follows from the definition of Ch$(\mathbb{P})$%
. }

{\small \textit{Claim 2.} Con$(\mathbb{P})=\varnothing $. To see this, take
any $S\in \mathfrak{X}$ and $x\in S,$ and let $I$ stand for the interval $(%
\mathbb{P}^{\ast }(x,S),\min_{w\in S}\mathbb{P}^{\ast }(x,\{x,w\})].$
Suppose $I\neq \varnothing $. Then, $\mathbb{P}^{\ast }(x,S)<1,$ so, since $%
\theta _{1}>\frac{1}{2},$ there exists a $y\in S$ such that $u_{1}(y)=\max
u_{1}(S)>u_{1}(x).$ If there is a $z\in S$ with $u_{1}(y)>u_{1}(z)>u_{1}(x),$
the the hypothesis of the proposition entails $u_{i}(y)>u_{i}(x)$ for all $%
i\in N,$ so the right endpoint of $I$ is $0,$ contradicting the nonemptiness
of $I$. Assume, then, $u_{1}(y)>u_{1}(x)>u_{1}(z)$ for all $z\in S\backslash
\{x,y\}.$ Then, by hypothesis, $u_{i}(y)>u_{i}(z)$ for all $i\in N$ and $%
z\in S\backslash \{y\}.$ It is then easy to check that $\mathbb{P}(y, \{x,
y\})=\mathbb{P}(y, S)$ and $\mathbb{P}(x, \{x, y\})=\mathbb{P}(x, S)$, which
means $I=\varnothing $, a contradiction. It then follows from the definition
of the Condorcet set of $\mathbb{P}$ that Con$(\mathbb{P})=\varnothing .$ }

{\small \textit{Claim 3.} ST$(\mathbb{P})=\varnothing $. To see this, take
any $x,y,z\in X$ with $\mathbb{P}^{\ast }(y,\{x,y\})<1$ and $\mathbb{P}%
^{\ast }(z,\{y,z\})<1.$ Since $\theta _{1}>\frac{1}{2}$, then, $%
u_{1}(x)>u_{1}(y)>u_{1}(z),$ so by hypothesis, $u_{i}(x)>u_{i}(z)$ for all $%
i\in N,$ whence $\mathbb{P}^{\ast }(z,\{x,z\})=0.$ Our claim thus follows
from the definition of ST$(\mathbb{P})$. }

{\small Claims 1-3 jointly imply $\Lambda (\mathbb{P})=\varnothing $$,$
completing our proof. $\blacksquare $ }

\bigskip

{\small \noindent \textsc{Proof of Proposition 5.5.}\footnote{%
This proof was suggested to us by Gil Riella.}\ Take any $u:X\rightarrow
(0,\infty )$ and a partial order $\succcurlyeq $ on $X,$ and let $\mathbb{P}$
be a proper 2-stage Luce model induced by $(u,\succcurlyeq ).$ Fix an
arbitrary $\lambda $ in $(0,1],$ and define the partial order $\succcurlyeq
_{\lambda }$ on $X$ by $x\succcurlyeq _{\lambda }y$ iff either $x=y$ or $%
\lambda u(x)>u(y).$ It is readily checked that $\succcurlyeq _{\lambda }$ is 
$\succcurlyeq $-transitive, that is,%
\begin{equation}
\text{either }x\succ _{\lambda }y\succcurlyeq z\text{ or }x\succcurlyeq
y\succ _{\lambda }z\text{ implies }x\succ _{\lambda }z  \label{pf}
\end{equation}%
for every $x,y,z\in X,$ and%
\begin{equation}
C_{\mathbb{P},\lambda }(S)=\text{{\footnotesize \textbf{MAX}}}%
(S,\succcurlyeq _{\lambda })\cap \text{{\footnotesize \textbf{MAX}}}%
(S,\succcurlyeq )  \label{ppf}
\end{equation}%
for every $S\in \mathfrak{X}.$ As an easy consequence of this
characterization, we find that $C_{\mathbb{P},\lambda }$ satisfies the
Chernoff and Condorcet axioms. To verify the No-Cycle axiom, take any $%
x,y,z\in X$ such that $\{x\}=C_{\mathbb{P},\lambda }\{x,y\}$ and $\{y\}=C_{%
\mathbb{P},\lambda }\{y,z\}$. In view of (\ref{ppf}), the first of these
equations implies that either $x\succ y$ or $x\succ _{\lambda }y$, and the
second implies that either $y\succ z$ or $y\succ _{\lambda }z$. Using (\ref%
{pf}), we then find that $x\succ _{\lambda }z,$ that is, {\footnotesize 
\textbf{MAX}}$(\{x,z\},\succcurlyeq _{\lambda })=\{x\}.$ Besides $x\in $ 
{\footnotesize \textbf{MAX}}$(\{x,z\},\succcurlyeq ),$ because, otherwise, $%
z\succ x$, so properness of $\mathbb{P}$ entails $u(z)>u(x)>\lambda u(x),$
so not $x\succ _{\lambda }z,$ a contradiction. It then follows from (\ref%
{ppf}) that $\{z\}=C_{\mathbb{P},\lambda }\{x,z\}.$ Conclusion: $C_{\mathbb{P%
},\lambda }$ satisfies the No-Cycle axiom. Thus, by Theorem 2.2, $C_{\mathbb{%
P},\lambda }$ is rational. $\blacksquare $ }

\bigskip

\noindent \textbf{References}

\smallskip

\noindent Afriat, S., On a system of inequalities in demand analysis: An
extension of the classical method, \textit{Inter. Econ. Rev.} 14 (1973),
460-472.

\smallskip

\noindent Aguiar, V., Random categorization and bounded rationality, \textit{%
Econ. Lett.} 159 (2017), 46-52.

\smallskip

\noindent Aguiar, V., and R. Serrano, Slutsky matrix norms: The size,
classification, and comparative statics of bounded rationality, \textit{J.
Econ. Theory} 172 (2017), 163-201.

\smallskip

\noindent Agranov, M., and P. Ortoleva, Stochastic choice and preferences for randomization, 
\textit{J. Polit. Econ}, 125 (2017), 40-68.

\smallskip

\noindent Agranov, M., and P. Ortoleva, Range of randomization, 
\textit{Rev. Econ. Stat}, forthcoming, 2023.

%von Neumann and Morgenstern (1947), Aumann (1962), Bewley (1986), and Schmeidler (1989)

\smallskip

\noindent Apesteguia, J., and M. Ballester, A measure of rationality and
welfare, \textit{J. Polit. Econ.} 123 (2015), 1278-1310.

\smallskip

\noindent Apesteguia, J., and M. Ballester, Separating predicted randomness
from residual behavior, \textit{J. Europ. Econ. Assoc.} 19 (2021), 1041-1076.

\smallskip

\noindent Apesteguia, J., M. Ballester, and J. Lu, Single-crossing random
utility models, \textit{Econometrica} 85 (2017), 661-674.

\smallskip

\noindent Ahumada, A., and L. Ulku, Luce rule with limited consideration, 
\textit{Math. Soc. Sci.} 93 (2018), 52-56.

\smallskip

\noindent Aumann, R.J., Utility theory without the completeness axiom, 
\textit{Econometrica} 30 (1962), 445-462.

\smallskip

\noindent Balakrishnan, N., E. A. Ok, and P. Ortoleva, Inferential choice
theory, \textit{mimeo}, Princeton Univ., 2021.

\smallskip

\noindent Bewley, T.F., Knightian Decision Theory: Part I, Cowles Foundation
Discussion Paper No. 807, 1986.

\smallskip 

\noindent Buchanan, J., Social choice, democracy, and free markets, \textit{%
J. Polit. Econ.} 62 (1954), 114-123.

\smallskip

\noindent {Caradonna, P., The implications of experimental design for choice
data, \textit{mimeo}, Georgetown Univ., 2020.}

\smallskip

\noindent Cerreia-Vioglio, S., D. Dillenberger, P. Ortoleva, and G. Riella,
Deliberately stochastic, \textit{Amer. Econ. Rev,} 109 (2019),
2425-2445.

\smallskip

\noindent Choi, S., R. Fisman, D. Gale, and S. Kariv,
Consistency and heterogeneity of individual behavior under uncertainty, \textit{Amer. Econ. Rev,} 97 (2007),
1921-1938.

%\smallskip

%\noindent de Clippel, G., and K. Rozen, Which performs best? Comparing
%discrete choice models, \textit{mimeo}, Brown Univ., 2022.

\smallskip

\noindent de Clippel, G., and K. Rozen, Bounded rationality in choice: A
survey, \textit{J. Econ. Lit.} forthcoming, 2023.

\smallskip

\noindent Dean, M., and D. Martin, Measuring rationality with the minimum
cost of revealed preference relations, \textit{Rev. Econ. Stat.} 98 (2016),
524-534.

\smallskip

\noindent Echenique, F., S. Lee, and M. Shum, The money pump as a measure of
revealed preference violations, \textit{J. Polit. Econ.} 119 (2019),
1201-1223.

\smallskip

\noindent Echenique, F., and K. Saito, General Luce model, \textit{Econ.
Theory} 68 (2019), 811-826.

\smallskip

\noindent Echenique, F., K. Saito, and G. Tserenjigmid, The
perception-adjusted Luce model, \textit{Math. Soc. Sci.} 93 (2018), 67-76.

\smallskip

\noindent Eliaz, K.\ and E. A. Ok, Indifference or indecisiveness:
Choice-theoretic foundations of incomplete preferences, \textit{Games Econ.
Behav.}\ 56 (2006), 61-86.

\smallskip

\noindent {Falmagne, J., A representation theorem for finite random scale
systems,\ }\textit{J. Math. Psych.} 18 (1978), 52-72.

\smallskip

\noindent {Feldman, P., and J. Rehbeck, Revealing a Preference for Mixtures:
An Experimental Study of Risk,\ }\textit{Quant. Econ.} 13 (2022), 761-786.

\smallskip

\noindent {Fishburn, P., Choice probabilities and choice functions,\ }%
\textit{J. Math. Psych.} 18 (1978), 201-219.

\smallskip

\noindent Fishburn, P., Induced binary probabilities and the linear ordering
polytope: A status report, \textit{Math. Soc. Sci. }23 (1992), 67-80.

\smallskip

\noindent Fudenberg, D., R. Iijima, and T. Strzalecki, Stochastic choice and
revealed perturbed utility, \textit{Econometrica} 83 (2015), 2371-2409.

\smallskip

\noindent {Gerasimou, G., Indecisiveness, undesirability and overload
revealed through choice deferral\ }\textit{Econ. J.} 128 (2018), 2450-2479.

\smallskip

\noindent Gul, F., and W. Pesendorfer, Random choice as behavioral
optimization, \textit{Econometrica} 82 (2014), 1873-1912.

\smallskip

\noindent Gul, F., P. Natenzon, and W. Pesendorfer, Random choice as
behavioral optimization, \textit{Econometrica} 82 (2014), 1873-1912.

\smallskip

\noindent Harless, D., and C. Camerer, The predictive utility of generalized
expected utility theories, \textit{Econometrica} 62 (1994), 1251-1289.

\smallskip

\noindent He, J., and P. Natenzon, Moderate utility, \textit{Amer. Econ. Rev: Insights}, forthcoming, 2023.

\smallskip

\noindent Hey, J., Does repetition improve consistency? \textit{Exp. Econ.}, 4 (2001), 5-54.

\smallskip

\noindent Hey, J., and C. Orme, Investigating generalizations of expected utility using experimental data, \textit{Econometrica}, 62 (1994), 1291-1326.

\smallskip

\noindent Horan, S., Stochastic semi-orders, \textit{J. Econ. Theory} 192
(2021), 1051-71.

\smallskip

\noindent Houtman, M., and J. Maks, Determining all maximal data subsets
consistent with revealed preference, \textit{Kwantitatieve Methoden} 6
(1985), 89-104.

\smallskip

\noindent Kitamura, Y., and J. Stoye, Nonparametric analysis of random
utility models, \textit{Econometrica} 86 (2018), 1883-1909.

\smallskip

\noindent Luce, D., \textit{Individual Choice Behavior: A Theoretical
Analysis},\textit{\ }New York, Wiley, 1959.

\smallskip

\noindent Machina, M., Stochastic choice functions generated from
deterministic preferences over lotteries, \textit{Econ. J.} 95 (1985),
575-594.

\smallskip

\noindent Manzini, P., and M. Mariotti, Sequentially rationalizable choice,\ 
\textit{Amer. Econ. Rev.} 97 (2007), 1824-1839.

\smallskip

\noindent Manzini, P., and M. Mariotti, Stochastic choice and consideration
sets,\ \textit{Econometrica} 82 (2014), 1153-1176.

\smallskip

\noindent Manzini, P., and M. Mariotti, Dual random utility maximisation,\ 
\textit{J. Econ. Theory} 177 (2018), 162-182.

\smallskip

\noindent Masatlioglu, Y., D. Nakajima, and E. Ozbay, Revealed attention, 
\textit{Amer. Econ. Rev.} 102 (2012), 2183-2205.

\smallskip

\noindent McFadden, D., Revealed stochastic preference: A synthesis, \textit{%
Economic Theory} 26 (2005), 245-264.

\smallskip

\noindent Natenzon, P., Random choice and learning, \textit{J. Polit. Econ. }%
127 (2019), 419-457.

\smallskip

\noindent Nielsen, K., and L. Rigotti, Revealed Incomplete Preferences, 
\textit{mimeo}, Caltech., 2023.

\smallskip

\noindent Ok, E. A., and G. Tserenjigmid, Indifference, indecisiveness,
experimentation and stochastic choice, \textit{Theoretical Econ. }17 (2022),
378-413.

\smallskip

\noindent Regenwetter, M., J. Dana, and C. Davis-Stober, Transitivity of
preferences, \textit{Psychol. Rev.} 118 (2011), 42-56.

\smallskip

\noindent Ribeiro, M., and G. Riella, Regular preorders and behavioral
indifference, \textit{Theory Dec.} 82 (2017), 1-12.

\smallskip

\noindent Ribeiro, M., Comparative rationality, \textit{mimeo}, New York
Univ., 2021.

\smallskip

\noindent Schmeidler, D., Subjective probability and expected utility
without additivity, \textit{Econometrica} 57 (1989), 571-587.

\smallskip

\noindent Sprumont, Y., Regular random choice and the triangle inequalities, 
\textit{J. Math. Psych.} 110 (2022), 102710.

\smallskip

\noindent Strzalecki, T., Stochastic choice theory, \textit{mimeo}, Harvard
Univ., 2023.

\smallskip

\noindent Tversky, A., Intransitivity of preferences, \textit{Psychol. Rev.}
76 (1969), 31-48.

\smallskip

\noindent Tversky, A., Elimination by aspects: A theory of choice, \textit{%
Psychol. Rev}. 79 (1972), 281-299.

\smallskip

\noindent Varian, H., Goodness-of-fit in optimizing models, \textit{J.
Econometrics} 46 (1990), 125-140.

\smallskip 

\noindent von Neumann, J., and O. Morgenstern, \textit{Theory of Games and
Economic Behavior}, Princeton Univ. Press, 1947.

%TCIMACRO{\TeXButton{TeX field}{\end{small}}}%
%BeginExpansion
\end{small}%
%EndExpansion

\end{document}